\documentclass[longauth]{aa}
\usepackage{graphicx}
\usepackage{natbib}
\usepackage{scalerel}
\usepackage{multirow}

\usepackage[table]{xcolor}

\usepackage{txfonts}
\usepackage[pdfencoding=auto,psdextra]{hyperref}
\hypersetup{
    colorlinks=true,
    linkcolor=blue,
    filecolor=magenta,      
    urlcolor=blue,
    citecolor=blue
}
\makeatletter
\renewcommand*\aa@pageof{, page \thepage{} of \pageref*{LastPage}}
\makeatother

\usepackage[utf8]{inputenc}

\usepackage[switch, modulo]{lineno}
\renewcommand{\linenumbers}[0]{}

\usepackage{euclid}

\begin{document}
%
%

\title{The PAU Survey and \Euclid: Analysing photometric redshifts from realistically simulated narrow-band photometry with Flagship 2\thanks{This paper is published on behalf of the Euclid Consortium.}}

   
\newcommand{\orcid}[1]{} 

\author{A.~Wittje\orcid{0000-0002-8173-3438}\thanks{\email{awitt@astro.ruhr-uni-bochum.de}}\inst{\ref{aff1}}
\and H.~Hildebrandt\orcid{0000-0002-9814-3338}\inst{\ref{aff1}}
\and D.~Navarro-Giron\'{e}s\orcid{0000-0003-0507-372X}\inst{\ref{aff2},\ref{aff3},\ref{aff4}}
\and E.~J.~Gonzalez\orcid{0000-0002-0226-9893}\inst{\ref{aff5},\ref{aff6}}
\and J.~L.~van den Busch\inst{\ref{aff1}}
\and E.~Gaztanaga\orcid{0000-0001-9632-0815}\inst{\ref{aff3},\ref{aff4},\ref{aff7}}
\and M.~Eriksen\orcid{0000-0003-0601-0990}\inst{\ref{aff8},\ref{aff9}}
\and A.~H.~Wright\orcid{0000-0001-7363-7932}\inst{\ref{aff1}}
\and R.~Casas\orcid{0000-0002-8165-5601}\inst{\ref{aff4},\ref{aff3}}
\and J.~Carretero\orcid{0000-0002-3130-0204}\inst{\ref{aff10},\ref{aff9}}
\and F.~J.~Castander\orcid{0000-0001-7316-4573}\inst{\ref{aff3},\ref{aff4}}
\and E.~Fernandez\inst{\ref{aff8}}
\and P.~Fosalba\orcid{0000-0002-1510-5214}\inst{\ref{aff4},\ref{aff3}}
\and J.~Garc\'ia-Bellido\orcid{0000-0002-9370-8360}\inst{\ref{aff11}}
\and H.~Hoekstra\orcid{0000-0002-0641-3231}\inst{\ref{aff2}}
\and R.~Miquel\orcid{0000-0002-6610-4836}\inst{\ref{aff8},\ref{aff12}}
\and C.~Padilla\orcid{0000-0001-7951-0166}\inst{\ref{aff8}}
\and P.~Renard\orcid{0000-0002-5953-4491}\inst{\ref{aff3},\ref{aff13}}
\and E.~Sanchez\orcid{0000-0002-9646-8198}\inst{\ref{aff10}}
\and I.~Sevilla-Noarbe\orcid{0000-0002-1831-1953}\inst{\ref{aff10}}
\and P.~Tallada-Cresp\'{i}\orcid{0000-0002-1336-8328}\inst{\ref{aff10},\ref{aff9}}
\and B.~Altieri\orcid{0000-0003-3936-0284}\inst{\ref{aff14}}
\and S.~Andreon\orcid{0000-0002-2041-8784}\inst{\ref{aff15}}
\and N.~Auricchio\orcid{0000-0003-4444-8651}\inst{\ref{aff16}}
\and C.~Baccigalupi\orcid{0000-0002-8211-1630}\inst{\ref{aff17},\ref{aff18},\ref{aff19},\ref{aff20}}
\and M.~Baldi\orcid{0000-0003-4145-1943}\inst{\ref{aff21},\ref{aff16},\ref{aff22}}
\and S.~Bardelli\orcid{0000-0002-8900-0298}\inst{\ref{aff16}}
\and P.~Battaglia\orcid{0000-0002-7337-5909}\inst{\ref{aff16}}
\and A.~Biviano\orcid{0000-0002-0857-0732}\inst{\ref{aff18},\ref{aff17}}
\and M.~Bolzonella\orcid{0000-0003-3278-4607}\inst{\ref{aff16}}
\and E.~Branchini\orcid{0000-0002-0808-6908}\inst{\ref{aff23},\ref{aff24},\ref{aff15}}
\and M.~Brescia\orcid{0000-0001-9506-5680}\inst{\ref{aff25},\ref{aff26}}
\and S.~Camera\orcid{0000-0003-3399-3574}\inst{\ref{aff27},\ref{aff28},\ref{aff29}}
\and G.~Ca\~nas-Herrera\orcid{0000-0003-2796-2149}\inst{\ref{aff2}}
\and V.~Capobianco\orcid{0000-0002-3309-7692}\inst{\ref{aff29}}
\and C.~Carbone\orcid{0000-0003-0125-3563}\inst{\ref{aff30}}
\and V.~F.~Cardone\inst{\ref{aff31},\ref{aff32}}
\and M.~Castellano\orcid{0000-0001-9875-8263}\inst{\ref{aff31}}
\and G.~Castignani\orcid{0000-0001-6831-0687}\inst{\ref{aff16}}
\and S.~Cavuoti\orcid{0000-0002-3787-4196}\inst{\ref{aff26},\ref{aff33}}
\and A.~Cimatti\inst{\ref{aff34}}
\and C.~Colodro-Conde\inst{\ref{aff35}}
\and G.~Congedo\orcid{0000-0003-2508-0046}\inst{\ref{aff36}}
\and L.~Conversi\orcid{0000-0002-6710-8476}\inst{\ref{aff37},\ref{aff14}}
\and Y.~Copin\orcid{0000-0002-5317-7518}\inst{\ref{aff38}}
\and F.~Courbin\orcid{0000-0003-0758-6510}\inst{\ref{aff39},\ref{aff12},\ref{aff13}}
\and H.~M.~Courtois\orcid{0000-0003-0509-1776}\inst{\ref{aff40}}
\and M.~Crocce\orcid{0000-0002-9745-6228}\inst{\ref{aff3},\ref{aff4}}
\and H.~Degaudenzi\orcid{0000-0002-5887-6799}\inst{\ref{aff41}}
\and G.~De~Lucia\orcid{0000-0002-6220-9104}\inst{\ref{aff18}}
\and H.~Dole\orcid{0000-0002-9767-3839}\inst{\ref{aff42}}
\and F.~Dubath\orcid{0000-0002-6533-2810}\inst{\ref{aff41}}
\and C.~A.~J.~Duncan\orcid{0009-0003-3573-0791}\inst{\ref{aff36}}
\and X.~Dupac\inst{\ref{aff14}}
\and S.~Escoffier\orcid{0000-0002-2847-7498}\inst{\ref{aff43}}
\and M.~Farina\orcid{0000-0002-3089-7846}\inst{\ref{aff44}}
\and R.~Farinelli\inst{\ref{aff16}}
\and S.~Farrens\orcid{0000-0002-9594-9387}\inst{\ref{aff45}}
\and F.~Faustini\orcid{0000-0001-6274-5145}\inst{\ref{aff31},\ref{aff46}}
\and S.~Ferriol\inst{\ref{aff38}}
\and F.~Finelli\orcid{0000-0002-6694-3269}\inst{\ref{aff16},\ref{aff47}}
\and S.~Fotopoulou\orcid{0000-0002-9686-254X}\inst{\ref{aff48}}
\and N.~Fourmanoit\orcid{0009-0005-6816-6925}\inst{\ref{aff43}}
\and M.~Frailis\orcid{0000-0002-7400-2135}\inst{\ref{aff18}}
\and M.~Fumana\orcid{0000-0001-6787-5950}\inst{\ref{aff30}}
\and S.~Galeotta\orcid{0000-0002-3748-5115}\inst{\ref{aff18}}
\and K.~George\orcid{0000-0002-1734-8455}\inst{\ref{aff49}}
\and B.~Gillis\orcid{0000-0002-4478-1270}\inst{\ref{aff36}}
\and C.~Giocoli\orcid{0000-0002-9590-7961}\inst{\ref{aff16},\ref{aff22}}
\and J.~Gracia-Carpio\orcid{0000-0003-4689-3134}\inst{\ref{aff50}}
\and A.~Grazian\orcid{0000-0002-5688-0663}\inst{\ref{aff51}}
\and F.~Grupp\inst{\ref{aff50},\ref{aff52}}
\and S.~V.~H.~Haugan\orcid{0000-0001-9648-7260}\inst{\ref{aff53}}
\and W.~Holmes\orcid{0009-0007-8554-4646}\inst{\ref{aff54}}
\and F.~Hormuth\inst{\ref{aff55}}
\and A.~Hornstrup\orcid{0000-0002-3363-0936}\inst{\ref{aff56},\ref{aff57}}
\and K.~Jahnke\orcid{0000-0003-3804-2137}\inst{\ref{aff58}}
\and M.~Jhabvala\inst{\ref{aff59}}
\and B.~Joachimi\orcid{0000-0001-7494-1303}\inst{\ref{aff60}}
\and S.~Kermiche\orcid{0000-0002-0302-5735}\inst{\ref{aff43}}
\and A.~Kiessling\orcid{0000-0002-2590-1273}\inst{\ref{aff54}}
\and M.~Kilbinger\orcid{0000-0001-9513-7138}\inst{\ref{aff45}}
\and B.~Kubik\orcid{0009-0006-5823-4880}\inst{\ref{aff38}}
\and M.~K\"ummel\orcid{0000-0003-2791-2117}\inst{\ref{aff52}}
\and H.~Kurki-Suonio\orcid{0000-0002-4618-3063}\inst{\ref{aff61},\ref{aff62}}
\and A.~M.~C.~Le~Brun\orcid{0000-0002-0936-4594}\inst{\ref{aff63}}
\and S.~Ligori\orcid{0000-0003-4172-4606}\inst{\ref{aff29}}
\and P.~B.~Lilje\orcid{0000-0003-4324-7794}\inst{\ref{aff53}}
\and V.~Lindholm\orcid{0000-0003-2317-5471}\inst{\ref{aff61},\ref{aff62}}
\and I.~Lloro\orcid{0000-0001-5966-1434}\inst{\ref{aff64}}
\and G.~Mainetti\orcid{0000-0003-2384-2377}\inst{\ref{aff65}}
\and O.~Mansutti\orcid{0000-0001-5758-4658}\inst{\ref{aff18}}
\and O.~Marggraf\orcid{0000-0001-7242-3852}\inst{\ref{aff66}}
\and M.~Martinelli\orcid{0000-0002-6943-7732}\inst{\ref{aff31},\ref{aff32}}
\and N.~Martinet\orcid{0000-0003-2786-7790}\inst{\ref{aff67}}
\and F.~Marulli\orcid{0000-0002-8850-0303}\inst{\ref{aff68},\ref{aff16},\ref{aff22}}
\and R.~J.~Massey\orcid{0000-0002-6085-3780}\inst{\ref{aff69}}
\and E.~Medinaceli\orcid{0000-0002-4040-7783}\inst{\ref{aff16}}
\and S.~Mei\orcid{0000-0002-2849-559X}\inst{\ref{aff70},\ref{aff71}}
\and M.~Melchior\inst{\ref{aff72}}
\and M.~Meneghetti\orcid{0000-0003-1225-7084}\inst{\ref{aff16},\ref{aff22}}
\and E.~Merlin\orcid{0000-0001-6870-8900}\inst{\ref{aff31}}
\and G.~Meylan\orcid{0000-0001-6503-0209}\inst{\ref{aff73}}
\and A.~Mora\orcid{0000-0002-1922-8529}\inst{\ref{aff74}}
\and M.~Moresco\orcid{0000-0002-7616-7136}\inst{\ref{aff68},\ref{aff16}}
\and L.~Moscardini\orcid{0000-0002-3473-6716}\inst{\ref{aff68},\ref{aff16},\ref{aff22}}
\and R.~Nakajima\orcid{0009-0009-1213-7040}\inst{\ref{aff66}}
\and C.~Neissner\orcid{0000-0001-8524-4968}\inst{\ref{aff8},\ref{aff9}}
\and S.-M.~Niemi\orcid{0009-0005-0247-0086}\inst{\ref{aff75}}
\and S.~Paltani\orcid{0000-0002-8108-9179}\inst{\ref{aff41}}
\and F.~Pasian\orcid{0000-0002-4869-3227}\inst{\ref{aff18}}
\and K.~Pedersen\inst{\ref{aff76}}
\and V.~Pettorino\orcid{0000-0002-4203-9320}\inst{\ref{aff75}}
\and A.~Pezzotta\orcid{0000-0003-0726-2268}\inst{\ref{aff15}}
\and S.~Pires\orcid{0000-0002-0249-2104}\inst{\ref{aff45}}
\and G.~Polenta\orcid{0000-0003-4067-9196}\inst{\ref{aff46}}
\and M.~Poncet\inst{\ref{aff77}}
\and L.~A.~Popa\inst{\ref{aff78}}
\and L.~Pozzetti\orcid{0000-0001-7085-0412}\inst{\ref{aff16}}
\and F.~Raison\orcid{0000-0002-7819-6918}\inst{\ref{aff50}}
\and A.~Renzi\orcid{0000-0001-9856-1970}\inst{\ref{aff79},\ref{aff80},\ref{aff16}}
\and J.~Rhodes\orcid{0000-0002-4485-8549}\inst{\ref{aff54}}
\and G.~Riccio\inst{\ref{aff26}}
\and E.~Romelli\orcid{0000-0003-3069-9222}\inst{\ref{aff18}}
\and M.~Roncarelli\orcid{0000-0001-9587-7822}\inst{\ref{aff16}}
\and R.~Saglia\orcid{0000-0003-0378-7032}\inst{\ref{aff52},\ref{aff50}}
\and Z.~Sakr\orcid{0000-0002-4823-3757}\inst{\ref{aff11},\ref{aff81},\ref{aff82}}
\and A.~G.~S\'anchez\orcid{0000-0003-1198-831X}\inst{\ref{aff50}}
\and D.~Sapone\orcid{0000-0001-7089-4503}\inst{\ref{aff83}}
\and B.~Sartoris\orcid{0000-0003-1337-5269}\inst{\ref{aff52},\ref{aff18}}
\and P.~Schneider\orcid{0000-0001-8561-2679}\inst{\ref{aff66}}
\and T.~Schrabback\orcid{0000-0002-6987-7834}\inst{\ref{aff84}}
\and A.~Secroun\orcid{0000-0003-0505-3710}\inst{\ref{aff43}}
\and E.~Sefusatti\orcid{0000-0003-0473-1567}\inst{\ref{aff18},\ref{aff17},\ref{aff19}}
\and E.~Sihvola\orcid{0000-0003-1804-7715}\inst{\ref{aff85}}
\and P.~Simon\inst{\ref{aff66}}
\and C.~Sirignano\orcid{0000-0002-0995-7146}\inst{\ref{aff79},\ref{aff80}}
\and G.~Sirri\orcid{0000-0003-2626-2853}\inst{\ref{aff22}}
\and A.~N.~Taylor\inst{\ref{aff36}}
\and H.~I.~Teplitz\orcid{0000-0002-7064-5424}\inst{\ref{aff86}}
\and I.~Tereno\orcid{0000-0002-4537-6218}\inst{\ref{aff87},\ref{aff88}}
\and N.~Tessore\orcid{0000-0002-9696-7931}\inst{\ref{aff89}}
\and S.~Toft\orcid{0000-0003-3631-7176}\inst{\ref{aff90},\ref{aff91}}
\and R.~Toledo-Moreo\orcid{0000-0002-2997-4859}\inst{\ref{aff92},\ref{aff93}}
\and F.~Torradeflot\orcid{0000-0003-1160-1517}\inst{\ref{aff9},\ref{aff10}}
\and I.~Tutusaus\orcid{0000-0002-3199-0399}\inst{\ref{aff3},\ref{aff4},\ref{aff81}}
\and J.~Valiviita\orcid{0000-0001-6225-3693}\inst{\ref{aff61},\ref{aff62}}
\and T.~Vassallo\orcid{0000-0001-6512-6358}\inst{\ref{aff18},\ref{aff49}}
\and G.~Verdoes~Kleijn\orcid{0000-0001-5803-2580}\inst{\ref{aff94}}
\and A.~Veropalumbo\orcid{0000-0003-2387-1194}\inst{\ref{aff15},\ref{aff24},\ref{aff23}}
\and Y.~Wang\orcid{0000-0002-4749-2984}\inst{\ref{aff95}}
\and J.~Weller\orcid{0000-0002-8282-2010}\inst{\ref{aff52},\ref{aff50}}
\and G.~Zamorani\orcid{0000-0002-2318-301X}\inst{\ref{aff16}}
\and F.~M.~Zerbi\orcid{0000-0002-9996-973X}\inst{\ref{aff15}}
\and E.~Zucca\orcid{0000-0002-5845-8132}\inst{\ref{aff16}}
\and A.~Montoro\orcid{0000-0003-4730-8590}\inst{\ref{aff3},\ref{aff4}}
\and M.~Sereno\orcid{0000-0003-0302-0325}\inst{\ref{aff16},\ref{aff22}}}
										   
\institute{Ruhr University Bochum, Faculty of Physics and Astronomy, Astronomical Institute (AIRUB), German Centre for Cosmological Lensing (GCCL), 44780 Bochum, Germany\label{aff1}
\and
Leiden Observatory, Leiden University, Einsteinweg 55, 2333 CC Leiden, The Netherlands\label{aff2}
\and
Institute of Space Sciences (ICE, CSIC), Campus UAB, Carrer de Can Magrans, s/n, 08193 Barcelona, Spain\label{aff3}
\and
Institut d'Estudis Espacials de Catalunya (IEEC),  Edifici RDIT, Campus UPC, 08860 Castelldefels, Barcelona, Spain\label{aff4}
\and
Departament de F\'{\i}sica, Universitat Aut\`onoma de Barcelona, 08193 Bellaterra (Barcelona), Spain\label{aff5}
\and
Instituto de Astronomia Teorica y Experimental (IATE-CONICET), Laprida 854, X5000BGR, C\'ordoba, Argentina\label{aff6}
\and
Institute of Cosmology and Gravitation, University of Portsmouth, Portsmouth PO1 3FX, UK\label{aff7}
\and
Institut de F\'{i}sica d'Altes Energies (IFAE), The Barcelona Institute of Science and Technology, Campus UAB, 08193 Bellaterra (Barcelona), Spain\label{aff8}
\and
Port d'Informaci\'{o} Cient\'{i}fica, Campus UAB, C. Albareda s/n, 08193 Bellaterra (Barcelona), Spain\label{aff9}
\and
Centro de Investigaciones Energ\'eticas, Medioambientales y Tecnol\'ogicas (CIEMAT), Avenida Complutense 40, 28040 Madrid, Spain\label{aff10}
\and
Instituto de F\'isica Te\'orica UAM-CSIC, Campus de Cantoblanco, 28049 Madrid, Spain\label{aff11}
\and
Instituci\'o Catalana de Recerca i Estudis Avan\c{c}ats (ICREA), Passeig de Llu\'{\i}s Companys 23, 08010 Barcelona, Spain\label{aff12}
\and
Institut de Ciencies de l'Espai (IEEC-CSIC), Campus UAB, Carrer de Can Magrans, s/n Cerdanyola del Vall\'es, 08193 Barcelona, Spain\label{aff13}
\and
ESAC/ESA, Camino Bajo del Castillo, s/n., Urb. Villafranca del Castillo, 28692 Villanueva de la Ca\~nada, Madrid, Spain\label{aff14}
\and
INAF-Osservatorio Astronomico di Brera, Via Brera 28, 20122 Milano, Italy\label{aff15}
\and
INAF-Osservatorio di Astrofisica e Scienza dello Spazio di Bologna, Via Piero Gobetti 93/3, 40129 Bologna, Italy\label{aff16}
\and
IFPU, Institute for Fundamental Physics of the Universe, via Beirut 2, 34151 Trieste, Italy\label{aff17}
\and
INAF-Osservatorio Astronomico di Trieste, Via G. B. Tiepolo 11, 34143 Trieste, Italy\label{aff18}
\and
INFN, Sezione di Trieste, Via Valerio 2, 34127 Trieste TS, Italy\label{aff19}
\and
SISSA, International School for Advanced Studies, Via Bonomea 265, 34136 Trieste TS, Italy\label{aff20}
\and
Dipartimento di Fisica e Astronomia, Universit\`a di Bologna, Via Gobetti 93/2, 40129 Bologna, Italy\label{aff21}
\and
INFN-Sezione di Bologna, Viale Berti Pichat 6/2, 40127 Bologna, Italy\label{aff22}
\and
Dipartimento di Fisica, Universit\`a di Genova, Via Dodecaneso 33, 16146, Genova, Italy\label{aff23}
\and
INFN-Sezione di Genova, Via Dodecaneso 33, 16146, Genova, Italy\label{aff24}
\and
Department of Physics "E. Pancini", University Federico II, Via Cinthia 6, 80126, Napoli, Italy\label{aff25}
\and
INAF-Osservatorio Astronomico di Capodimonte, Via Moiariello 16, 80131 Napoli, Italy\label{aff26}
\and
Dipartimento di Fisica, Universit\`a degli Studi di Torino, Via P. Giuria 1, 10125 Torino, Italy\label{aff27}
\and
INFN-Sezione di Torino, Via P. Giuria 1, 10125 Torino, Italy\label{aff28}
\and
INAF-Osservatorio Astrofisico di Torino, Via Osservatorio 20, 10025 Pino Torinese (TO), Italy\label{aff29}
\and
INAF-IASF Milano, Via Alfonso Corti 12, 20133 Milano, Italy\label{aff30}
\and
INAF-Osservatorio Astronomico di Roma, Via Frascati 33, 00078 Monteporzio Catone, Italy\label{aff31}
\and
INFN-Sezione di Roma, Piazzale Aldo Moro, 2 - c/o Dipartimento di Fisica, Edificio G. Marconi, 00185 Roma, Italy\label{aff32}
\and
INFN section of Naples, Via Cinthia 6, 80126, Napoli, Italy\label{aff33}
\and
Dipartimento di Fisica e Astronomia "Augusto Righi" - Alma Mater Studiorum Universit\`a di Bologna, Viale Berti Pichat 6/2, 40127 Bologna, Italy\label{aff34}
\and
Instituto de Astrof\'{\i}sica de Canarias, E-38205 La Laguna, Tenerife, Spain\label{aff35}
\and
Institute for Astronomy, University of Edinburgh, Royal Observatory, Blackford Hill, Edinburgh EH9 3HJ, UK\label{aff36}
\and
European Space Agency/ESRIN, Largo Galileo Galilei 1, 00044 Frascati, Roma, Italy\label{aff37}
\and
Universit\'e Claude Bernard Lyon 1, CNRS/IN2P3, IP2I Lyon, UMR 5822, Villeurbanne, F-69100, France\label{aff38}
\and
Institut de Ci\`{e}ncies del Cosmos (ICCUB), Universitat de Barcelona (IEEC-UB), Mart\'{i} i Franqu\`{e}s 1, 08028 Barcelona, Spain\label{aff39}
\and
UCB Lyon 1, CNRS/IN2P3, IUF, IP2I Lyon, 4 rue Enrico Fermi, 69622 Villeurbanne, France\label{aff40}
\and
Department of Astronomy, University of Geneva, ch. d'Ecogia 16, 1290 Versoix, Switzerland\label{aff41}
\and
Universit\'e Paris-Saclay, CNRS, Institut d'astrophysique spatiale, 91405, Orsay, France\label{aff42}
\and
Aix-Marseille Universit\'e, CNRS/IN2P3, CPPM, Marseille, France\label{aff43}
\and
INAF-Istituto di Astrofisica e Planetologia Spaziali, via del Fosso del Cavaliere, 100, 00100 Roma, Italy\label{aff44}
\and
Universit\'e Paris-Saclay, Universit\'e Paris Cit\'e, CEA, CNRS, AIM, 91191, Gif-sur-Yvette, France\label{aff45}
\and
Space Science Data Center, Italian Space Agency, via del Politecnico snc, 00133 Roma, Italy\label{aff46}
\and
INFN-Bologna, Via Irnerio 46, 40126 Bologna, Italy\label{aff47}
\and
School of Physics, HH Wills Physics Laboratory, University of Bristol, Tyndall Avenue, Bristol, BS8 1TL, UK\label{aff48}
\and
University Observatory, LMU Faculty of Physics, Scheinerstr.~1, 81679 Munich, Germany\label{aff49}
\and
Max Planck Institute for Extraterrestrial Physics, Giessenbachstr. 1, 85748 Garching, Germany\label{aff50}
\and
INAF-Osservatorio Astronomico di Padova, Via dell'Osservatorio 5, 35122 Padova, Italy\label{aff51}
\and
Universit\"ats-Sternwarte M\"unchen, Fakult\"at f\"ur Physik, Ludwig-Maximilians-Universit\"at M\"unchen, Scheinerstr.~1, 81679 M\"unchen, Germany\label{aff52}
\and
Institute of Theoretical Astrophysics, University of Oslo, P.O. Box 1029 Blindern, 0315 Oslo, Norway\label{aff53}
\and
Jet Propulsion Laboratory, California Institute of Technology, 4800 Oak Grove Drive, Pasadena, CA, 91109, USA\label{aff54}
\and
Felix Hormuth Engineering, Goethestr. 17, 69181 Leimen, Germany\label{aff55}
\and
Technical University of Denmark, Elektrovej 327, 2800 Kgs. Lyngby, Denmark\label{aff56}
\and
Cosmic Dawn Center (DAWN), Denmark\label{aff57}
\and
Max-Planck-Institut f\"ur Astronomie, K\"onigstuhl 17, 69117 Heidelberg, Germany\label{aff58}
\and
NASA Goddard Space Flight Center, Greenbelt, MD 20771, USA\label{aff59}
\and
Department of Physics and Astronomy, University College London, Gower Street, London WC1E 6BT, UK\label{aff60}
\and
Department of Physics, P.O. Box 64, University of Helsinki, 00014 Helsinki, Finland\label{aff61}
\and
Helsinki Institute of Physics, Gustaf H{\"a}llstr{\"o}min katu 2, University of Helsinki, 00014 Helsinki, Finland\label{aff62}
\and
Laboratoire d'etude de l'Univers et des phenomenes eXtremes, Observatoire de Paris, Universit\'e PSL, Sorbonne Universit\'e, CNRS, 92190 Meudon, France\label{aff63}
\and
SKAO, Jodrell Bank, Lower Withington, Macclesfield SK11 9FT, UK\label{aff64}
\and
Centre de Calcul de l'IN2P3/CNRS, 21 avenue Pierre de Coubertin 69627 Villeurbanne Cedex, France\label{aff65}
\and
Universit\"at Bonn, Argelander-Institut f\"ur Astronomie, Auf dem H\"ugel 71, 53121 Bonn, Germany\label{aff66}
\and
Aix-Marseille Universit\'e, CNRS, CNES, LAM, Marseille, France\label{aff67}
\and
Dipartimento di Fisica e Astronomia "Augusto Righi" - Alma Mater Studiorum Universit\`a di Bologna, via Piero Gobetti 93/2, 40129 Bologna, Italy\label{aff68}
\and
Department of Physics, Institute for Computational Cosmology, Durham University, South Road, Durham, DH1 3LE, UK\label{aff69}
\and
Universit\'e Paris Cit\'e, CNRS, Astroparticule et Cosmologie, 75013 Paris, France\label{aff70}
\and
CNRS-UCB International Research Laboratory, Centre Pierre Bin\'etruy, IRL2007, CPB-IN2P3, Berkeley, USA\label{aff71}
\and
University of Applied Sciences and Arts of Northwestern Switzerland, School of Engineering, 5210 Windisch, Switzerland\label{aff72}
\and
Institute of Physics, Laboratory of Astrophysics, Ecole Polytechnique F\'ed\'erale de Lausanne (EPFL), Observatoire de Sauverny, 1290 Versoix, Switzerland\label{aff73}
\and
Telespazio UK S.L. for European Space Agency (ESA), Camino bajo del Castillo, s/n, Urbanizacion Villafranca del Castillo, Villanueva de la Ca\~nada, 28692 Madrid, Spain\label{aff74}
\and
European Space Agency/ESTEC, Keplerlaan 1, 2201 AZ Noordwijk, The Netherlands\label{aff75}
\and
DARK, Niels Bohr Institute, University of Copenhagen, Jagtvej 155, 2200 Copenhagen, Denmark\label{aff76}
\and
Centre National d'Etudes Spatiales -- Centre spatial de Toulouse, 18 avenue Edouard Belin, 31401 Toulouse Cedex 9, France\label{aff77}
\and
Institute of Space Science, Str. Atomistilor, nr. 409 M\u{a}gurele, Ilfov, 077125, Romania\label{aff78}
\and
Dipartimento di Fisica e Astronomia "G. Galilei", Universit\`a di Padova, Via Marzolo 8, 35131 Padova, Italy\label{aff79}
\and
INFN-Padova, Via Marzolo 8, 35131 Padova, Italy\label{aff80}
\and
Institut de Recherche en Astrophysique et Plan\'etologie (IRAP), Universit\'e de Toulouse, CNRS, UPS, CNES, 14 Av. Edouard Belin, 31400 Toulouse, France\label{aff81}
\and
Universit\'e St Joseph; Faculty of Sciences, Beirut, Lebanon\label{aff82}
\and
Departamento de F\'isica, FCFM, Universidad de Chile, Blanco Encalada 2008, Santiago, Chile\label{aff83}
\and
Universit\"at Innsbruck, Institut f\"ur Astro- und Teilchenphysik, Technikerstr. 25/8, 6020 Innsbruck, Austria\label{aff84}
\and
Department of Physics and Helsinki Institute of Physics, Gustaf H\"allstr\"omin katu 2, University of Helsinki, 00014 Helsinki, Finland\label{aff85}
\and
Infrared Processing and Analysis Center, California Institute of Technology, Pasadena, CA 91125, USA\label{aff86}
\and
Departamento de F\'isica, Faculdade de Ci\^encias, Universidade de Lisboa, Edif\'icio C8, Campo Grande, PT1749-016 Lisboa, Portugal\label{aff87}
\and
Instituto de Astrof\'isica e Ci\^encias do Espa\c{c}o, Faculdade de Ci\^encias, Universidade de Lisboa, Tapada da Ajuda, 1349-018 Lisboa, Portugal\label{aff88}
\and
Mullard Space Science Laboratory, University College London, Holmbury St Mary, Dorking, Surrey RH5 6NT, UK\label{aff89}
\and
Cosmic Dawn Center (DAWN)\label{aff90}
\and
Niels Bohr Institute, University of Copenhagen, Jagtvej 128, 2200 Copenhagen, Denmark\label{aff91}
\and
Universidad Polit\'ecnica de Cartagena, Departamento de Electr\'onica y Tecnolog\'ia de Computadoras,  Plaza del Hospital 1, 30202 Cartagena, Spain\label{aff92}
\and
European University of Technology EUt+, European Union\label{aff93}
\and
Kapteyn Astronomical Institute, University of Groningen, PO Box 800, 9700 AV Groningen, The Netherlands\label{aff94}
\and
Caltech/IPAC, 1200 E. California Blvd., Pasadena, CA 91125, USA\label{aff95}}    

%
%
\abstract{The study of the large-scale structure of the Universe and the distribution of galaxies has been greatly advanced by the application of various types of photometric redshift estimation techniques. 
The  Physics of the Accelerating Universe Survey (PAUS) is a state-of-the-art imaging survey, designed to provide high-quality photometric data in 40 optical narrow-band filters. 
In this paper, we present an in-depth analysis of the photometric redshifts obtained from PAUS using realistic simulations.
Based on the \Euclid Flagship 2 catalogue, we generate PAUS-like simulations with realistic photometric noise and target selection. The mock catalogue is used to explore the accuracy and precision of photometric redshifts of PAUS, investigating the impact of noise and depth, and combinations of broad- and narrow-band data.
The simulations reproduce the observed PAUS performance well, with a robust photometric redshift scatter of $\sigma_{\sfont{NMAD}}/(1+z)\!=\!0.011$ for a mock subsample resembling the spectroscopic validation sample of the data.
For the complete PAUS-like sample, which can only be evaluated in the simulation, we find $\sigma_{\sfont{NMAD}}/(1+z)\!=\!0.028$ for $i_{\rm AB}\!\lesssim\!23$ and outlier fractions below 10\%.
Forecasts indicate that including \Euclid photometry in a joint analysis could further reduce scatter and outlier fractions, especially for faint galaxy populations.
The hypothetical case of deeper PAUS observations over a limited sky area, forecasting the potential for future narrow- or medium-band surveys, in combination with \Euclid's deep photometry, achieves scatter as low as $\sigma_{\sfont{NMAD}}/(1+z)\!=\!0.011$ and outlier fractions below 4\% for $i_{\rm AB}\!<\!23.5$.
    }
%
%
%
%
   \titlerunning{The PAU Survey and \Euclid\/: Analysing photometric redshifts from Flagship 2 narrow-band photometry}
   \authorrunning{A. Wittje et al.}
   
   \maketitle
%
%
%
%

\defcitealias{navarro-gironesPAUSurveyPhotometric2023}{N24}

\section{\label{sc:Intro}Introduction }
Understanding the distribution of galaxies across time is essential for studying structure formation, galaxy evolution, and the fundamental properties of the Universe. A key requirement for such studies is the determination of accurate distances to galaxies. While spectroscopic redshifts (spec-$z$s) provide the most precise measurements, their acquisition is observationally expensive and limited to rather bright samples of galaxies \citep[e.g.,][]{lefevreVIMOSVLTDeep2005}. As a result, alternative approaches are required to estimate redshifts efficiently across the large galaxy samples produced by modern surveys.

Photometric redshifts, also referred to as photo-$z$s, provide such an approach by inferring redshift estimates from galaxy fluxes measured through multiple photometric bands \citep[e.g.,][]{benitezBayesianPhotometricRedshift2000,ilbertAccuratePhotometricRedshifts2006,hildebrandtCARSCFHTLSArchiveResearchSurvey2009}. Cosmological surveys such as the Kilo-Degree Survey \citep[KiDS;][]{kuijkenGravitationalLensingAnalysis2015}, the Dark Energy Survey \citep[DES;][]{abbottDarkEnergySurvey2018}, the Hyper Suprime-Cam Subaru Strategic Program \citep[HSC-SSP;][]{aiharaFirstDataRelease2018}, the \Euclid Wide Survey and Deep Fields \citep{Laureijs11, EuclidSkyOverview}, and the Vera C. Rubin Observatory Legacy Survey of Space and Time \citep[LSST;][]{ivezicLSSTScienceDrivers2019}, among others, demonstrate the power of multi-band imaging for redshift estimation. These surveys provide photometric redshifts over large cosmological volumes, enabling a broad range of applications, including large-scale structure mapping, studies of galaxy evolution, and constraints on cosmological parameters \citep[e.g.,][]{wrightKiDSLegacyCosmologicalConstraints2025,descollaborationDarkEnergySurvey2022,dalalHyperSuprimeCamYear2023}.

However, broad-band photo-$z$s are limited by several factors that affect both accuracy and reliability. The filter coverage, specifically meaning the wavelength range and placement of the available filters, determines whether key spectral features can be identified, while the quality and depth of the photometric data further impact precision. The availability of representative spectroscopic calibration samples is essential, but often incomplete or biased \citep[e.g.,][]{cunhaSpectroscopicFailuresPhotometric2014,newmanSpectroscopicNeedsImaging2015}. 
While the performance also depends on the different photo-$z$ methods \citep{salvatoManyFlavoursPhotometric2018}, the broad width of the filters mostly limits the accuracy by causing broad scatter and outliers due to degeneracies \citep{benitezBayesianPhotometricRedshift2000}.

To mitigate some of these limitations, several surveys have adopted narrower filter systems, albeit at the expense of survey area, galaxy density and/or depth compared to broad-band photometric surveys. Projects such as the Classifying Objects by Medium-Band Observations in 17 Filters survey (COMBO-17; \citealt{wolfCOMBO17SurveyEvolution2003}), the Advanced Large Homogeneous Area Medium Band Redshift Astronomical Survey (ALHAMBRA; \citealt{molinoALHAMBRASurveyBayesian2014}), the Cosmic Evolution Survey (COSMOS; \citealt{laigleCOSMOS2015CatalogExploring2016}), the Physics of the Accelerating Universe Survey (PAUS; \citealt{martiPrecisePhotometricRedshifts2014}), the Javalambre Physics of the Accelerating Universe Survey (J-PAS; \citealt{benitezJPASJavalambrePhysicsAccelerated2014,bonoliMiniJPASSurveyPreview2021}), the Javalambre-Photometric Local Universe Survey (J-PLUS; \citealt{cenarroJPLUSJavalambrePhotometric2019}), and the Southern Photometric Local Universe Survey (S-PLUS; \citealt{mendesdeoliveiraSouthernPhotometricLocal2019}), use narrow- or medium-band filters to measure the spectral energy distribution (SED) of distant sources with higher spectral resolution. As a result, the filter systems of these surveys provide an improvement in photo-$z$ precision by up to a factor of 4 compared to broad-band filters alone \citep{molinoAssessingPhotometricRedshift2020}, effectively bridging the gap between traditional photometric and spectroscopic redshift techniques.

In this study, we focus on PAUS, which utilises 40 narrow-band filters spanning the optical range. 
In various studies, PAUS has demonstrated its ability to obtain high-quality photometric redshifts in the COSMOS field and in wider fields using both template-fitting methods \citep{eriksenPAUSurveyEarly2019,navarro-gironesPAUSurveyPhotometric2023} and machine-learning approaches \citep{eriksenPAUSurveyPhotometric2020,daza-perillaPAUSurveyEnhancing2025}, making it a valuable resource for astrophysical and cosmological analyses \citep[e.g.,][]{johnstonPAUSurveyIntrinsic2021,renardPAUSurveyMeasurements2022, gonzalezPAUSurveyClose2023,navarro-gironesPAUSurveyMeasuring2025}. 
Nevertheless, the PAUS photo-$z$ performance at faint magnitudes ($i_{\rm AB} \gtrsim 22.5$) is not yet well analysed, since spectroscopic reference data are scarce in this regime.

Given the limited observational constraints, cosmological simulations play a central role in testing the accuracy of photometric redshift estimates for faint galaxies. Synthetic galaxy catalogues constructed from large-volume simulations allow photo-$z$ techniques to be tested under well-defined conditions and provide insights into systematic errors that cannot be disentangled from real data alone. 

In this work, we generate realistic PAUS-like mock catalogues based on the Euclid Flagship 2 simulation \citep{EuclidSkyFlagship} in order to benchmark photo-$z$ performance. This simulation has been developed to support the \Euclid mission, producing galaxy mocks with realistic clustering, photometric properties, and redshift distributions. By incorporating PAUS-like photometry for Flagship 2 galaxies, one can directly assess photo-$z$ precision and systematics across magnitude ranges, which are hardly accessible to spectroscopy. The main advantage of this simulation is that it provides complete galaxy samples with SEDs, enabling the computation of fluxes in multiple bands and making it well suited for our analysis.

We test the PAUS-specific photo-$z$ algorithm \texttt{BCNZ} \citep{eriksenPAUSurveyEarly2019} across different noise regimes, evaluate systematic uncertainties at the faint end, and compare mock results with real PAUS observations. 
Additionally, in the simulation, we can also explore potential improvements of the photo-$z$ statistics from combining PAUS with broad-band photometry from deep surveys such as \Euclid, as well as the hypothetical setup in which we increase the depth of PAUS flux measurements.

The structure of this paper is as follows. Section\,\ref{sc:Data} describes the PAUS data, the \Euclid mission, and the Flagship 2 simulation. Section\,\ref{sc:Mocks} outlines the methodology used to construct PAUS-like mocks with realistic noise levels. Section\,\ref{sc:Photo-z} presents the estimation and performance analysis of photo-$z$s. In Sect.\,\ref{sc:Results}, we compare the resulting photo-$z$ catalogues of the mocks with the PAUS data, and investigate extensions with \Euclid and a deeper version of PAUS in Sect.\,\ref{sec:deeper_paus}. Section\,\ref{sc:Conclusion} summarises our conclusions.

\section{\label{sc:Data}Data and simulation}

\subsection{PAUS data}
PAUS \citep[][]{martiPrecisePhotometricRedshifts2014,tonelloPAUSurveyOperation2019} is a photometric galaxy survey with 40 narrow-band images. The optical camera \citep[PAUCam,][]{padillaPhysicsAcceleratingUniverse2019} at the William Herschel Telescope (WHT) located on the Observatorio del Roque de los Muchachos (La Palma, Canary Islands, Spain) observed about 50\,deg$^2$ of the sky to $i_{\rm AB}\,\simeq\,23$.  
The 40 optical filters have a width of 13\,nm (full width at half maximum), are equally spaced (10\,nm apart from each other), and together cover a wavelength range from 450\,nm to 850\,nm. 
These narrow-band images from PAUS are combined with deep broad-band photometry from COSMOS \citep{laigleCOSMOS2015CatalogExploring2016}, the Canada-France-Hawaii Telescope Lensing Survey \citep[CFHTLenS,][]{heymansCFHTLenSCanadaFranceHawaiiTelescope2012}, or KiDS \citep[][]{jongKiloDegreeSurvey2013}. 

The galaxy fluxes are measured by the pipeline called \texttt{MEMBA} \citep{tonelloPAUSurveyOperation2019}. The galaxy positions in the PAUS fields are determined using the galaxy positions of the broad-band surveys, and the narrow-band fluxes are obtained using forced aperture photometry on the narrow-band images \citep{serranoPAUSurveyNarrowband2023}. The sizes of the elliptical apertures for the narrow bands are chosen to collect 62.5\% of a galaxy's total flux.
In the end, \texttt{MEMBA} provides fluxes that are background subtracted, scaled with the image zero points and combined into coadded fluxes using a weighted mean \citep{castanderNarrowBandPhotometry,eriksenPAUSurveyEarly2019}.

PAUS has observations in the CFHTLenS wide fields W1, W3, and W4 \citep{heymansCFHTLenSCanadaFranceHawaiiTelescope2012}, as well as part of the GAMA G09 field \citep{driverGalaxyMassAssembly2012}, and the COSMOS field. 
The COSMOS field \citep[][]{laigleCOSMOS2015CatalogExploring2016} was used as a calibration field since many photometric observations are publicly available, spanning from ultraviolet to far-infrared wavelengths. The PAUS calibration field, with an area of about 2\,deg$^2$, was extensively used to test photometric redshift measurements \citep{eriksenPAUSurveyEarly2019, alarconPAUSurveyImproved2021}.
In \citet[][hereafter N24]{navarro-gironesPAUSurveyPhotometric2023}, the photometric redshifts of the PAUS measurements in the W1, W3, and G09 fields were analysed, and these are used to construct the mock catalogues presented in this work. The W4 and COSMOS fields are not considered since they do not have much area with (overlapping) PAUS imaging and, therefore, there are not many photometric redshifts there. 

PAUS conducted observations in all 40 narrow bands for a total of 43\,deg$^{2}$ of the northern sky. Due to missing observations, more targets are observed with fewer than 40 narrow bands. \citetalias{navarro-gironesPAUSurveyPhotometric2023} found that robust photometric redshifts can be computed with 30 narrow bands or more, which results in a PAUS area of 51\,deg$^{2}$ containing 1.8 million objects. The PAUS W1 and W3 fields cover 12.04\,deg$^2$ and 22.64\,deg$^2$, respectively. The photo-$z$s are computed using the additional $ugriz$ broad-band data from CFHTLenS \citep{hildebrandtCFHTLenSImprovingQuality2012}.
The PAUS G09 field covers 15.7\,deg$^2$ and overlaps with the KiDS $ugriZYHK_{\rm s}$ broad-band data \citep{kuijkenFourthDataRelease2019}. The advantage here is the additional near-infrared bands, which are useful for computing photo-$z$s, since they extend the baseline wavelength range for the fitting of SED templates.

Beyond its role in improving photometric redshift estimates, PAUS data have supported a broad range of scientific applications, including measurements of the D4000 spectral break \citep{renardPAUSurveyMeasurements2022}, studies of close galaxy pairs and their mean halo masses \citep{gonzalezPAUSurveyClose2023}, comparisons with semi-analytical galaxy formation models \citep{manzoniPAUSurveyNew2024}, analysing the quasar luminosity functions at $2.7 < z < 5.3$ \citep{quasarluminosityfunctionsTorralba-Torregrosa}, and estimating the $i$-band galaxy luminosity function \citep{koonkorGalaxyIBandLuminosity}.
The quality of the photo-$z$s enabled the analysis of intrinsic alignments in the cosmic web \citep{johnstonPAUSurveyIntrinsic2021,navarro-gironesPAUSurveyMeasuring2025} and the calibration of the weak lensing redshift distributions \citep{vandenbuschKiDS1000CosmicShear2022,mylesDarkEnergySurvey2021}, demonstrating the wide scientific reach of PAUS. 
These early science results demonstrate the unique potential for both astrophysical and cosmological studies, while also highlighting the importance of characterising photo-$z$ performance across different ranges of magnitude and redshift.

\subsection{\Euclid}

The \Euclid satellite is designed to explore the composition and evolution of the dark Universe \citep{EuclidSkyOverview}. Its primary objective is to create a map of the large-scale structure of the Universe by measuring the shapes and positions of billions of galaxies, covering more than a third of the sky. By studying the Universe's expansion and the formation of cosmic structures over the past 10 billion years, \Euclid aims to reveal more about the properties of dark energy, dark matter, and gravity \citep{amendolaCosmologyFundamentalPhysics2018,Blanchard-EP7}.
\Euclid will observe galaxies in four different filter bands, ranging from the optical (\IE) to the near-infrared ($\YE,\JE,$ and $\HE$), with wavelengths from 460\,nm to 2000\,nm \citep{Schirmer-EP18}, allowing scientists to study properties and characteristics of distant galaxies.
\Euclid will measure their shapes \citep{EP-Congedo}, which are affected by shear, magnification, and intrinsic alignments caused by the presence of dark matter along the line of sight. These tiny distortions will be analysed statistically to reconstruct the distribution and evolution of dark matter with extreme precision.

Moreover, \Euclid will measure the redshifts of galaxies. For 25\,million emission-line galaxies over
the redshift range $0.9 < z < 1.8$, \Euclid will directly measure their redshifts through slitless spectroscopy at near-infrared wavelengths. Additionally, for 1.5\,billion galaxies, \Euclid will deliver accurate shape measurements. For these sources, photometric redshifts are estimated through combining \Euclid's photometry with the optical data from ground-based telescopes \citep{EuclidSkyOverview, Scaramella-EP1}. The complementary surveys at the end of the mission will be UNIONS \citep{gwynUNIONSUltravioletNearInfrared2025} and LSST \citep{lsstdarkenergysciencecollaborationLargeSynopticSurvey2012}, covering the northern and southern hemispheres, respectively. 

\subsection{Flagship 2}
The Euclid Flagship 2 Simulation \citep{ EuclidSkyFlagship} is a comprehensive cosmological simulation conducted by the Euclid Consortium.  
Flagship 2 aids in analysing and interpreting the data collected during the mission and in identifying and quantifying potential systematic effects in the analysis. 

We use the second version of the Flagship mock catalogue \citep{potterPKDGRAV3TrillionParticle2016}, which is a synthetic galaxy catalogue based on a dark-matter only simulation. It covers an area equivalent to an octant on the sky and contains 16\,000$^3$ particles in a simulation box with side length of 3600\,$h^{-1}$ Mpc up to a redshift of $z$\,=\,2.3. 
The galaxy mock is constructed similarly to the MICE2 simulation \citep{fosalbaMICEGrandChallenge2015}.
First, based on the dark matter distribution in Flagship 2, halo catalogues were built using a friends-of-friends halo finder \citep{crocceMICEGrandChallenge2015}. 
Haloes were populated with galaxies to describe the observed number densities and clustering using halo abundance and halo occupation models \citep{carreteroAlgorithmBuildMock2015}. 
Flagship 2 is a highly comprehensive catalogue that includes an extensive set of galaxy properties, such as positions, velocities, and photometry, as well as morphological and lensing parameters. This makes it an ideal tool for testing and validating analysis pipelines intended for application to observational data across multiple scientific use cases.
 
Flagship 2 (version 2.1.10\footnote{Available through Cosmohub \citep{CarreteroCosmohub2017,TALLADA2020100391} at \url{https://cosmohub.pic.es/catalogs/353}.}) provides a galaxy catalogue with fluxes (and estimates of their uncertainties) for over 60 different photometric filters from different telescopes and surveys. 
In addition, fluxes corresponding to the PAUS filter set were computed for specific sky patches of the Flagship 2 octant, of which we use 10\,deg$^2$ for the photo-$z$ estimation. For CFHTLenS and KiDS, a subset of the relevant photometric bands is included directly in Flagship 2, while additional bands were computed for the PAUS patches.
The CFHTLenS $ugriz$ filters are observed by MegaCam, and the KiDS set-up consists of the OmegaCam $ugri$ bands and the VISTA $ZYJHK_{\rm s}$ bands. 

Of the available options within Flagship 2, the true fluxes, $f^{\rm true}$, of the galaxies including the continuum and the emission lines, and including the internal attenuation by dust \citep[estimated with the Pozzetti dust model 3,][]{pozzettiExtremelyRedGalaxies2000} were chosen (i.e., those with the naming convention of \textit{survey\_filter}\_el\_model3\_ext). The extinction caused by the Milky Way is not included, given that PAUS fluxes are corrected for this effect.

\section{\label{sc:Mocks}Realistic simulations} \label{chap:mock_photometry}
For our analyses, we prepare mock catalogues that resemble the PAUS wide-field data \citepalias{navarro-gironesPAUSurveyPhotometric2023}.
In this section, we describe the process to obtain mock photometry with realistic noise for a large sample of galaxies and compute their photo-$z$s with the algorithm \texttt{BCNZ} \citep{eriksenPAUSurveyEarly2019}.

\begin{figure*}[h]
    \includegraphics[width=1\linewidth]{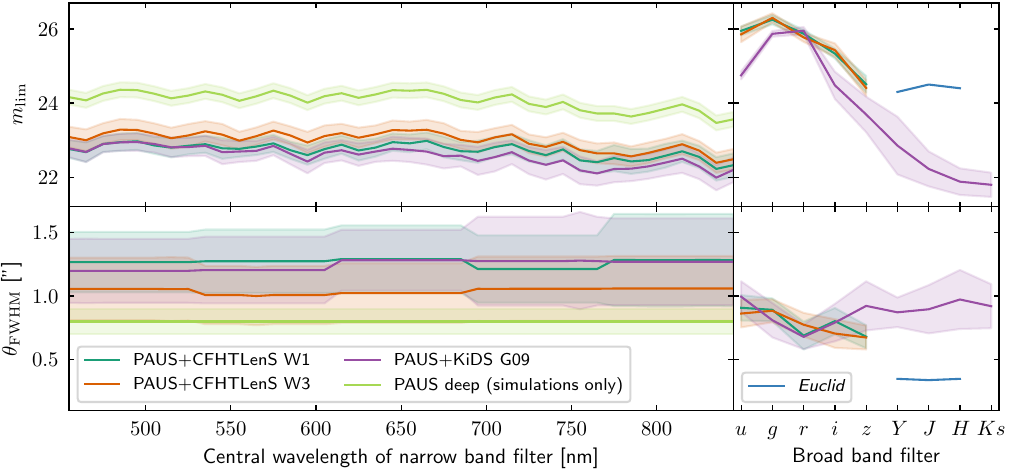}
    \caption{Mean 5\,$\sigma$ limiting magnitudes for point sources (top) and mean PSF FWHM values (bottom) for the PAUS narrow bands and overlapping broad-band surveys. Shaded areas indicate the standard error of the mean per PAUS field.  The broad-band values correspond to published CFHTLenS \citep{erbenCFHTLenSCanadaFranceHawaiiTelescope2013} and KiDS \citep{kuijkenFourthDataRelease2019} measurements. The values for a deeper PAUS configuration (explained in Sect.\,\ref{sec:deep_paus}) and \Euclid \citep{Scaramella-EP1, EuclidSkyNISP} are also shown. In \Euclid Q1, slightly differing limiting magnitudes were measured \citep{Q1-TP005} because of different aperture definitions. For consistency, we present the \Euclid 5\,$\sigma$ limiting magnitudes for point sources.    
    }
\label{fig:mag_limits_psf_data}
\end{figure*}
   
\subsection{Methodology for the noisy photometry of the PAUS mocks \label{sec:methodology_noise}}
To simulate the observed flux noise, we apply a method similar to the one presented in \citet{vandenbuschTestingKiDSCrosscorrelation2020}. 
In this modelling framework, the noise of a mock galaxy will depend on several parameters: its estimated projected size and the resulting aperture; its true flux; the seeing; and the limiting magnitude in a particular filter. 

To estimate the aperture that should be used to measure the flux of a galaxy, we will approximate an effective size of the mock galaxies with parameters given in the Flagship 2 mock, which are the half-light radius of a galaxy's bulge and disc, as well as the bulge-to-total flux ratio.
The projected size $R_{\mathrm{e},i}$ for each galaxy $i$ is then estimated from the weighted sum of the two components.
Based on the effective galaxy size and the mean seeing $\sigma_{\mathrm{\sfont{PSF}},x}$ per filter $x$, we create a mock aperture major axis
\begin{equation}
    A_{i,x} = \sqrt{R_{ \mathrm{e},i}^2+\sigma_{\mathrm{\sfont{PSF}},x}^2}\;, \label{eq:mock_aperture}
\end{equation}
and minor axis 
\begin{equation}
    B_{i,x} = \sqrt{\left[ \left(\frac{b}{ a}\right)_i R_{\mathrm{e},i}  \right]^2+\sigma_{\mathrm{\sfont{PSF}},x}^2}\;,
\end{equation}
with the projected axis ratio of the bulge $\left(b/a\right)_i$ where $b$ and $a$ are the semi-minor and semi-major axes, respectively, which are given for all Flagship 2 sources.

The seeing $\sigma_{\mathrm{\sfont{PSF}},x}$ measured in the real PAUS data is not a single fixed number because it varies with atmospheric conditions such as turbulence, humidity, and airmass. To mimic this behaviour in the simulations, we draw the seeing for each filter $x$ from a Gaussian distribution using the observed mean $\langle \sigma_{\mathrm{\sfont{PSF}},x} \rangle$ and its spread $\Delta \sigma_{\mathrm{\sfont{PSF}},x}$.

The knowledge of the simulated aperture can be converted into a weight for the flux error, which is one for point sources, and increases for larger objects. 
This signal-to-noise weight $W_{i,x}$ is defined as 
\begin{equation}
    W_{i,x} = \sqrt{A_{i,x}\, B_{i,x} \sigma_{\mathrm{\sfont{PSF}},x}^{-2}}\;. \label{eq:snw}
\end{equation}
Through weighting, we increase the flux error for galaxies with larger apertures relative to the seeing, thereby decreasing the signal-to-noise ratio (S/N) for extended objects. For a small, point-like source, this weight effectively gives a small flux error. 
The flux error $\Delta f_{i,x}$ is given as 
\begin{equation}
        5\,\Delta f_{i,x}= 10^{-0.4(m_{\mathrm{lim},x}-48.6)} \, W_{i,x}\;, \label{eq:flux_error}
\end{equation}
as per design using the 5\,$\sigma$ limiting magnitudes $m_{\mathrm{lim},x}$ per filter $x$.
Similar to the seeing values, the limiting magnitudes $m_{\mathrm{lim},x}$ are not treated as exact single values, but are sampled from Gaussian distributions with the mean and standard deviation estimated from the data.
The noisy flux $f^{\rm obs}_{i,x}$ is a flux realisation that is a random draw from a Gaussian distribution with mean $f^{\rm true}_{i,x}$ and a standard deviation of $\Delta f_{i,x}$.

\subsection{Seeing and limiting magnitudes from the data}
To complete the description of the noise model, Fig.\,\ref{fig:mag_limits_psf_data} presents the survey-specific inputs, which are the limiting magnitudes $m_{\rm lim}$ and the seeing as given by the full width at half maximum of the point spread function (PSF FWHM), $\theta_{\mathrm{\sfont{FWHM}}}$. These values are measured from the real PAUS data.

The depth of the PAUS observations is determined as the 5\,$\sigma$ limiting magnitudes $m_{\mathrm{lim}}$ for point sources, measured individually per filter and per PAUS field. 
Figure\,\ref{fig:mag_limits_psf_data} (top) shows the limiting magnitudes in the data, where the PAUS W3 field is slightly deeper in the narrow bands compared to W1 and G09, resulting from small variations in observing conditions. In the published photo-$z$ catalogues,\footnote{\url{https://cosmohub.pic.es/catalogs/319}} \citetalias{navarro-gironesPAUSurveyPhotometric2023} found no significant differences between the photo-$z$ quality of the W3 and W1 fields, but marginally better statistics than for the G09 field (Figure\,A3 in \citetalias{navarro-gironesPAUSurveyPhotometric2023}). This suggests, especially at the faint end of the observed galaxy sample, that the depth of the broad-band photometry is essential for the photo-$z$ estimation.
For the broad-band observations, we use the values of the limiting magnitudes from the literature as quoted in \citet{erbenCFHTLenSCanadaFranceHawaiiTelescope2013} and \citet{kuijkenFourthDataRelease2019} for CFHTLenS and KiDS, respectively.
CFHTLenS is approximately one magnitude deeper than KiDS in the $u$, $i$, and $z$ bands, and about half a magnitude deeper in the $g$ band. It is, however, shallower in the $r$ band, which is linked to the KiDS observing strategy. While the detection of objects in KiDS is performed in the $r$ band, in CFHTLenS the $i$ band was used. It is important to reproduce these survey-specific signatures in the mocks, since they propagate directly into the S/N distributions and ultimately into photo-$z$ accuracy. 

From the data, we also use the individual PSF FWHM values $\theta_{\mathrm{\sfont{FWHM}}}$ for each narrow band and for each PAUS field to reproduce the differences that can be present between the different PAUS wide fields.
Figure\,\ref{fig:mag_limits_psf_data} (bottom) shows the PSF FWHM as a function of the filter band and separated for each of the three PAUS wide fields. 
With the PAUcam, eight narrow bands were installed on the same filter tray and observed at the same time. This is visible in the resulting mean seeing, which is the same within a filter tray. The mean PSF FWHM values for the broad bands are taken from the archival data by CFHTLenS \citep{erbenCFHTLenSCanadaFranceHawaiiTelescope2013} and KiDS \citep{kuijkenFourthDataRelease2019}, specifically using the tiles that overlap with the PAUS wide fields. 
For the modelling framework presented in Sect.\,\ref{sec:methodology_noise}, the PSF values are converted from $\theta_{\rm FWHM}$ (see Fig.\,\ref{fig:mag_limits_psf_data}) to the standard deviation (assuming an approximate Gaussian distribution), obtaining $\sigma_{\mathrm{\sfont{PSF}}}$ used in Eqs.\,(\ref{eq:mock_aperture})--(\ref{eq:snw}).
Finally, we note that Fig. \ref{fig:mag_limits_psf_data} also includes the corresponding depths for \Euclid and PAUS deep, whose implications for our analysis will be discussed in more detail in Sect.\,\ref{sec:deeper_paus}.  

\subsection{PAUS apertures \label{sec:paus_apertures}}
In PAUS, the narrow-band photometry is measured using an elliptical, PSF-matched aperture that is defined to enclose 62.5\% of the total galaxy flux \citep{eriksenPAUSurveyEarly2019,serranoPAUSurveyNarrowband2023}. To ensure consistency when comparing to real data, we reproduce this aperture effect in the Flagship 2 mock catalogues by scaling the intrinsic total fluxes of each galaxy by a factor of 0.625. This approach assumes that the aperture fraction is approximately constant across the observed population, reflecting the PAUS strategy of fixing the flux fraction rather than the physical aperture size. Applying this scaling yields simulated narrow-band fluxes that are directly comparable to the measured PAUS fluxes, and avoids introducing systematic offsets arising from differing photometric measurement conventions.

\subsection{Simulating unobserved narrow-band fluxes}

As explained in \citetalias{navarro-gironesPAUSurveyPhotometric2023}, the PAUS wide fields do not have the full 40 narrow bands for the total area of 50\,deg$^2$. Incomplete coverage due to missing exposures is present, especially at the edges of the footprint. 
In the PAUS data, the `unobserved' fluxes are correlated for an individual object. If a pointing is not covered with a filter tray, then there is no photometry for the 8 narrow bands. The unobserved fluxes are also correlated for neighbouring galaxies within a tile. We can, however, compute reliable photo-$z$s for all the sources with observations of at least 30 narrow bands.  
We reproduce the distribution of the missing narrow-band observations by excluding narrow-band fluxes for a random subset of the Flagship 2 sources. For this Flagship 2 subset, we set the same narrow bands as `unobserved' as in the data. This ensures that the simulated catalogues capture the real-world variation in band coverage and do not artificially benefit from complete observations across the field.

\begin{figure*}[h]
    \centering
    \includegraphics[width=1\textwidth]{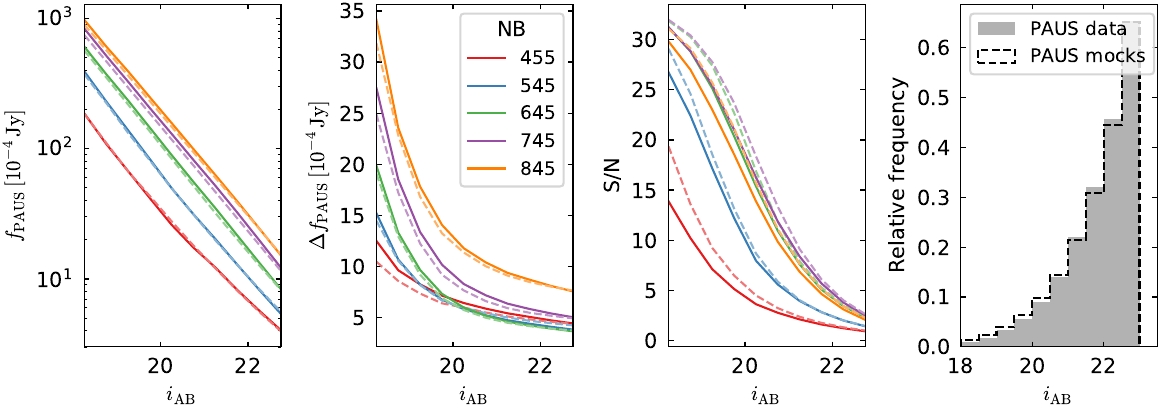}
    \caption{Comparison of narrow-band flux, flux error, S/N, and $i_{\rm AB}$ distribution of the PAUS data (solid) and simulated mocks (dashed). The colours in the first three panels indicate the central wavelength of each narrow-band filter (NB). 
    The brighter end shows slightly higher S/N in the mocks, but the fainter objects agree well with the data. 
    }
    \label{fig:flux_comp}
\end{figure*}

\subsection{\label{sec:compare_photometry}Comparing photometry to PAUS data}

With all the input values specific to each band and PAUS field, we obtain a set of mock photometry with a lot of realistic complexity. Equivalently to the data, we limit the sample to $i_{\rm AB} < 23$ for the PAUS fields overlapping CFHTLenS and $i_{\rm AB} < 23.1$ for the overlap with KiDS, to obtain comparable galaxy densities. The adjustment is due to the differences in the filter curves and depth in the broad-band surveys \citepalias{navarro-gironesPAUSurveyPhotometric2023}.
We can assess the accuracy of the mocks by comparing the flux and flux uncertainty distributions, as well as the S/N in all filter bands, with the real PAUS data.
In particular, we show the narrow-band fluxes in the same units as the PAUS wide-field data \citep{serranoPAUSurveyNarrowband2023}, which uses the magnitude definition according to 
\begin{equation}
    m_{\rm AB} = -2.5 \logten \left(\frac{f_{\rm \sfont{PAUS}}}{10^{-4}\,{\rm Jy}}\right) + 26\;.
\end{equation}

The fluxes and their uncertainties, especially for the narrow band, can be reproduced very well, as shown in Fig.\,\ref{fig:flux_comp}. One important quantity for the estimation of the photometric redshifts is the S/N in the individual bands, defined as 
S/N = $f^{\rm obs}$/$\Delta f^{\rm obs}$. Some disagreement is visible in a few bands at the brightest magnitudes in Fig.\,\ref{fig:flux_comp}. However, these objects represent only a negligible fraction of the sample, as evident from the magnitude distribution (right panel), and therefore do not affect the overall photometric-redshift performance.

The comparison for the CFHTLenS and KiDS broad-band photometry shows similarly good agreement between the data and the mocks in terms of magnitude, uncertainty, and S/N distributions.
For CFHTLenS, the mock broad-band S/N is comparable to or slightly lower than in the data, while for KiDS the agreement is generally good, although the mock $gri$ bands show somewhat higher S/N.
Overall, the broad-band photometric properties are reproduced sufficiently well for robust photo-$z$ validation.

\subsection{Mock spec-$z$ samples}
\label{sec:matched_sample}

To assess the photo-$z$ statistics of the data, a validation sample is used that consists of spectroscopic redshifts from surveys like the Sloan Digital Sky Survey DR16 \citep[SDSS,][]{ ahumadaSixteenthDataRelease2020}, GAMA DR3 \citep{baldryGalaxyMassAssembly2018}, the Vimos Public Redshift Survey \citep[VIPERS,][]{scodeggioVIMOSPublicExtragalactic2018}, the DEEP2 Galaxy Redshift Survey \citep{newmanDEEP2GalaxyRedshift2013}, and the VIMOS VLT Deep Survey \citep[VVDS,][]{lefevreVIMOSVLTDeep2013}, among others (with more details given in Appendix\,\ref{appendix:matched_samples}). Spectroscopic samples, however, are subject to a selection function, which results from various origins, including magnitude limits, fibre assignments for fibre spectrographs, targeting strategies, and noise in the fluxes, leading to failures in the redshift determination \citep{newmanSpectroscopicNeedsImaging2015}. 
The spectroscopic samples are therefore not a complete and representative subsample of the entire PAUS galaxy catalogue, since their selection function differs from that of the PAUS objects. 
In contrast, the galaxy mock provides the true redshift information for every object in the catalogue, and we can therefore assess the performance of the photo-$z$s in much more detail. 

To enable a fair and unbiased comparison between spectroscopic galaxies and those in the mocks, we construct mock subsamples that reproduce the selection functions applied to obtain the spec-$z$ validation samples.
This will also determine whether the validation sample for the PAUS sample yields reliable statistics for the whole data set. We recreate the validation sample in the mocks using a \textit{k}-nearest neighbours (\textit{k}NN) algorithm that was presented in \citet{wrightKiDSLegacyRedshiftDistributions2025}, where the authors use a multidimensional matching procedure to simulate calibration samples that closely mimic the real spectroscopic calibration data. We take a large `candidate' catalogue (from simulations) and a `target' catalogue (like the validation spec-$z$ sample of the data) and select, for each target source, the best-matching candidate in colour and magnitude space within a narrow true-redshift slice. The \textit{k}NN-matching proceeds by first restricting candidates to those whose true redshift lies near the target’s redshift, then computing the Euclidean distance in the feature space of photometric magnitudes (or colours) and selecting the candidate with the smallest distance (with optional duplication control).
This approach forces the simulated validation sample to replicate the joint distribution of redshift, magnitude, and colour of the real spectroscopic validation sample.

In this study, we tested the matching using only the broad-band information, using only the narrow-band information, or using all available bands together. Our tests showed the most robust results while only matching with the broad-band data. 
Including the narrow bands expanded the parameter space to more than 40 dimensions, significantly increasing the complexity of the optimisation. The parameter space for the matching becomes too sparsely populated to find a good match, and therefore leads to noisier results.
Moreover, the broad-band information alone ensures that the overall SED, combined with the redshift, gives a representative galaxy population that reproduces the spec-$z$ validation sample. This is also reasonable, since the initial sample selection for the spectroscopic samples (DEEP2 and VIPERS) was based on broad-band colours. 

\begin{figure}
    \centering
    \includegraphics[width=0.5\textwidth]{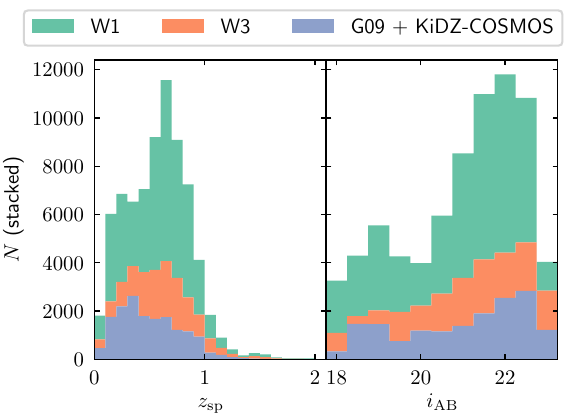}
    \caption{Redshift and $i_{\rm AB}$ distribution of the spectroscopic validation samples for W1, W3, and a combination of G09 and KiDZ-COSMOS, in a stacked histogram.
    Deeper KiDZ-COSMOS data supplement G09 to cover fainter magnitudes and higher redshifts. The mock samples by design mimic the redshift and magnitude distributions of these validation samples through the matching algorithm.}
    \label{fig:match_redshift_distr}
\end{figure}

Figure\,\ref{fig:match_redshift_distr} illustrates the spectroscopic data sets used for the photo-$z$ calibration of the PAUS data. The redshift distributions highlight the different depths of the spectroscopic campaigns: while W1 and W3 are well covered by VVDS and VIPERS, amongst others, the G09 field requires supplementation with deeper KiDZ-COSMOS spectroscopy \citep{wrightFifthDataRelease2024}. By construction, the matched mock samples replicate the magnitude and colour distributions of the validation data (for more details, see Appendix\,\ref{appendix:matched_samples}). 

\section{\label{sc:Photo-z}Photometric redshift methodology}
As shown in \citet{eriksenPAUSurveyEarly2019}, photometric redshifts for the 40 narrow bands of PAUS can be successfully calculated using the photo-$z$ algorithm \texttt{BCNZ}.
This is a template-fitting code designed to be used with the PAUS narrow-band photometry combined with external broad-band photometry. Even with the successful advance of machine-learning-based redshift estimation, template-fitting algorithms remain an essential component of photo-$z$ analyses. Their primary advantage lies in their independence from a complete spectroscopic training sample, which is often unavailable or biased against faint or high-redshift galaxies. Template-based methods can be readily applied to new surveys, are physically interpretable, and allow the propagation of photometric uncertainties in a statistically consistent way.

Contrary to other template-fitting codes like Bayesian Photometric Redshift  \citep[\texttt{BPZ},][]{benitezBayesianPhotometricRedshift2000}, \texttt{LePhare} \citep{ilbertAccuratePhotometricRedshifts2006}, and \texttt{EAZY} \citep{brammerEAZYFastPublic2008}, \texttt{BCNZ} additionally allows up to ten emission lines to be added. This is critical for leveraging the spectral resolution of the filter setup and for achieving the high precision in photo-$z$ required for the PAUS science goals. 

\subsection{\texttt{BCNZ} photo-$z$ algorithm}
\texttt{BCNZ} compares the observed flux with the fluxes of redshift-dependent model SEDs in each band. In addition to fitting linear combinations of templates, the algorithm simultaneously accounts for small photometric calibration differences by introducing per-band zero-point corrections and a global flux scaling between narrow- and broad-band data. The observed flux of a galaxy is a wavelength-dependent convolution of the emitted galaxy SED and the instrument's response. For the mocks, we simulated the noisy photometry of PAUS as described in Sect.\,\ref{chap:mock_photometry}. The template configurations follow the implementation presented in \citet{eriksenPAUSurveyEarly2019}. The \texttt{BCNZ} template set includes elliptical and spiral galaxy SEDs, starburst templates from the BC03 library \citep{bruzualStellarPopulationSynthesis2003}, and additional quiescent galaxy templates with exponentially declining star-formation histories, following \citet{ilbertGALAXYStellarMass2009} and \citet{laigleCOSMOS2015CatalogExploring2016}. For the starburst populations, dust attenuation is modelled using the \citet{calzettiDustContentOpacity2000} extinction law and two modified versions including a 2175 \AA\ bump, with $E(B-V)$ values ranging from 0.05 to 0.5. Elliptical and red spiral templates are used without extinction. 
A set of optical emission lines is included through additional templates. 
The emission-line templates adopt fixed relative line ratios, normalised to the [\ion{O}{ii}] flux, following the implementation of \citet{eriksenPAUSurveyEarly2019}. The adopted ratios are based on COSMOS2015 \citep{laigleCOSMOS2015CatalogExploring2016} and references therein.
The template configuration adopted here does not include dedicated AGN templates. While narrow-band surveys are in principle well suited to identifying such sources through their spectral features, specialised template sets and priors are generally required for reliable AGN photo-$z$ estimation \citep[e.g.,][]{salvatoDissectingPhotometricRedshift2009,hsuCANDELSMultiwavelengthCatalog2014}. For more details on the template sets, we refer to \citet{eriksenPAUSurveyEarly2019}. 

\texttt{BCNZ} uses a linear combination of templates $\alpha$ to fit the galaxy fluxes. 
The algorithm estimates the redshift probability distribution $p(z)$ for each galaxy through
\begin{equation}
    p(z) \propto \int_{\alpha_1 \geq 0} \diff \alpha_1 \,\, \dots \,\, \int_{\alpha_n \geq 0} \diff \alpha_n \,\, \exp\!\left(-0.5\, \chi^2[z,\alpha]\right)\,p_{\rm prior}(z,\alpha)\;,
\end{equation}
which schematically represents a Bayesian marginalisation over the non-negative template amplitudes and possible prior information on $z$ and $\alpha$. In practice, \texttt{BCNZ} does not adopt an explicit continuous Bayesian prior on galaxy type or redshift. Instead, the amplitudes are constrained to be positive, and the fit is carried out over a restricted set of allowed template combinations, which effectively acts as an implicit prior on the SED mixtures. Approximating the integral at each redshift by the maximum-likelihood solution (minimum $\chi^2$), the probability distribution can be written as
\begin{equation}
    p(z) \propto \exp\!\left[-0.5\, \chi_{\rm min}^2(z)\right]\;. \label{eq:probability_photoz}
\end{equation}

For \texttt{BCNZ}, the expression to minimise contains information about both the narrow and broad bands, abbreviated as NBs and BBs, respectively. Therefore the $\chi^2$ is defined as 
\begin{equation}
    \chi^2[z,\alpha] = \sum_{i,\rm NB} \left( \frac{f_i^{\rm obs} - l_i\,k\, f_i^{\rm model}}{\Delta f_i}\right)^2 + \sum_{i,\rm BB} \left( \frac{f_i^{\rm obs} - l_i\,f_i^{\rm model}}{\Delta f_i}\right)^2\,, \label{eq:minimisation}
\end{equation}
where for every band $i$, we have the observed flux $f_i^{\rm obs}$, its uncertainty $\Delta f_i$, a free scaling parameter $k$, and the `zero point' $l_i$. 
The standard zero points of the PAUS images were determined and corrected for before the coaddition of individual exposures. In the mocks, we do not have to correct for this effect. However, \texttt{BCNZ} performs a calibration step, where the zero points $l_i$ from Eq.\,(\ref{eq:minimisation}) are determined. These zero points describe the flux scaling between the observed (noisy) fluxes and the model fluxes of the template SEDs, which is estimated for each band individually.
Additionally, the external broad-band data use different apertures than the PAUS data. In \texttt{BCNZ}, the broad-band fluxes define the reference, while the PAUS narrow-band fluxes are rescaled to this system through the free factor $k$ in Eq.\,(\ref{eq:minimisation}). Physically, this factor accounts for the aperture mismatch between the PAUS narrow-band photometry and the external broad-band measurements.
For the mocks, the factor $k$ is constant, since we rescale the flux to replicate the PAUS apertures that encompass exactly 62.5\% of the total flux, as described in Sect.\,\ref{sec:paus_apertures}. For consistency with the treatment of real data, we keep $k$ as a free parameter in the minimisation, although in the mocks the true scaling is constant by construction.

The model flux $f_i^{\rm model}$ is defined as
\begin{equation}
    f_i^{\rm model}[z,\alpha] = \sum_{j=1}^n f_i^{j}(z) \, \alpha_j\;, \label{eq:combining_sed}
\end{equation}
with $f_i^{j}$ being the model flux of template $j$ in band $i$ with amplitude $\alpha_j$. The final template is therefore a linear combination
of multiple templates.
The probability distribution of the redshift of a galaxy can be approximated as 
\begin{equation}
    p(z) \propto \sum_{\mu} \exp\!\left[-0.5\, \chi_{\rm min, \alpha^{\mu}}^2(z)\right]\;, \label{eq:probability_photoz_approx}
\end{equation}
where the sum is over different sets of SEDs $\alpha^{\mu}$, with the approximation being the same as in Eq.\,(\ref{eq:probability_photoz}).

\texttt{BCNZ} performs the minimisation of $\chi^2$ (see Eq.\,\ref{eq:minimisation}) for each object in the PAUS sample. Knowing the fluxes of the narrow and broad bands and their uncertainties, the model fluxes $f_i^{\rm model}$ of combinations of templates can be fitted to the free parameter of interest, which is the photometric redshift $z_{\rm ph}$. 
As a result, the code determines a photometric redshift for every galaxy in the redshift range from 0.01 to 2 in steps of $\delta z$\,=\,0.001. 

\subsection{Statistical quality measures} \label{sec:statistics}

To determine the accuracy of the estimated photometric redshifts from PAUS data, they are compared to spectroscopic redshifts. For the mocks, we have the true redshift $z_{\rm true}$ of all the galaxies in the catalogue. The accuracy of the photometric redshift can be quantified with the photo-$z$ scatter, the outlier fraction, and the photo-$z$ bias.

The photo-$z$ error is calculated through comparing both redshift estimates, $z_{\rm ph}$ and $z_{\rm true}$, and is defined as 
\begin{equation}
    \Delta z = \frac{|z_{\rm ph}-z_{\rm true}|}{1+z_{\rm true}}\;.
\end{equation}
To measure the quality of the photometric redshifts, the scatter in photo-$z$ can be expressed with the normalised median absolute deviation (NMAD), which is a robust statistical measure of the spread of a distribution. It is defined as
\begin{equation} \label{eq:nmad}
\sigma_\sfont{NMAD} = 1.4826 \; \mathrm{median}\!\left( \, \left| \Delta z_i - \mathrm{median}(\Delta z) \right| \, \right)\;,
\end{equation} 
where the constant $1.4826$ makes $\sigma_\sfont{NMAD}$ consistent with the standard deviation for normally distributed data. However, unlike the standard deviation, $\sigma_\sfont{NMAD}$ is less sensitive to outliers.

To learn about the robustness of the photometric redshifts, it is useful to estimate the fraction of objects that have ‘catastrophic’ errors.
A galaxy is considered an outlier when the difference between its photometric and true redshift is much larger than the expected scatter. 
As in \citetalias{navarro-gironesPAUSurveyPhotometric2023}, we define a galaxy as an outlier, $\eta_{\sfont{0.1}}$, when the photo-$z$ error exceeds the limit of 
\begin{equation} \label{eq:outlier}
\Delta z > 0.1.
\end{equation}

Finally, the photo-$z$ bias refers to a systematic offset or discrepancy between the estimated redshift and the true redshift. By quantifying and minimising bias, the reliability of cosmological analysis can be improved. It is defined as 
\begin{equation} \label{eq:mu}
    \mu =\text{median}\left(\frac{z_{\rm ph}-z_{\rm true}}{1+z_{\rm true}}\right)\;.
\end{equation}
The photo-$z$ scatter given as $\sigma_\sfont{NMAD}$, the bias $\mu$, and the outlier fraction $\eta_{\sfont{0.1}}$, will be quantified as a function of the $i_{\rm AB}$ magnitude, the spectroscopic redshift, and the photometric redshift.

\subsection{Weighted photometric redshifts}
\label{sec:weighted_photo-z}

For bright objects, the much narrower PAUS filters yield an improvement in photo-$z$ precision, reducing the scatter $\sigma_\sfont{NMAD}$ by nearly an order of magnitude compared to the broad-band photo-$z$ estimates \citepalias{navarro-gironesPAUSurveyPhotometric2023}. For faint galaxies ($i_{\mathrm{AB}} > 22.5$), however, the PAUS narrow-band photometry becomes strongly noise-dominated, with S/N dropping to 1--3. This is because the narrow-band fluxes are limited by Poisson sky noise, while the broad-band data, taken from much deeper exposures and integrating flux over a wider wavelength range, retain relatively high S/N. Another reason could be the reduced effectiveness of the \texttt{BCNZ} fitting procedure when the input photometry becomes highly noise-dominated. As a result, the accuracy of \texttt{BCNZ} photo-$z$ estimates does not improve compared to broad-band-only methods for $i_{\mathrm{AB}} > 22.5$.

In contrast, photometric redshift algorithms for broad-band photometry like \texttt{BPZ} incorporate a strong, empirically calibrated Bayesian prior on the joint probability of redshift and spectral type as a function of apparent magnitude \citep{benitezBayesianPhotometricRedshift2000}. This prior downweights astrophysically implausible redshift–template combinations and regularises the solution when the photometric constraints are weak. At faint magnitudes, where \texttt{BCNZ} becomes highly degenerate, the absence of such a general astrophysical prior leads to an increased incidence of systematic outliers. In this regime, the \texttt{BPZ} prior effectively suppresses spurious secondary likelihood peaks and stabilises the redshift inference from the deeper broad-band data. 
Combining \texttt{BCNZ} with \texttt{BPZ} is therefore advantageous: \texttt{BCNZ} provides high spectral resolution at bright magnitudes, while \texttt{BPZ} supplies a robust, prior-driven constraint at the faint end where the narrow-band information becomes noise-dominated.

To exploit the complementary strengths of both approaches, we generate a weighted combination of the \texttt{BCNZ} and \texttt{BPZ} photo-$z$s. \citetalias{navarro-gironesPAUSurveyPhotometric2023} introduce a weighted photo-$z$ estimator as
\begin{equation}
    z_{ \sfont{\rm BCNZ}w} = \frac{z_{\rm \sfont{BCNZ}} \, w_\sfont{\rm BCNZ} + z_{\rm \sfont{BPZ}} \, w_\sfont{\rm BPZ}}{w_\sfont{BCNZ} +  w_\sfont{\rm BPZ}}\;,
    \label{eq:weighted_zp}
\end{equation}
where the weights $w$ were estimated from the photo-$z$ scatter as a function of $i_{\rm AB}$ magnitude, measured with either \texttt{BCNZ} or \texttt{BPZ}. In this work, we define them as $w$ = 1/$\sigma_{\sfont{NMAD},i_{\rm AB}}$. The \texttt{BPZ} estimates were found to be rather biased as a function of $i_{\rm AB}$ magnitude. Therefore, before applying Eq.\,(\ref{eq:weighted_zp}), we include a bias subtraction, also as a function of $i_{\rm AB}$ magnitude, to avoid propagating systematic offsets in \texttt{BPZ}.

This scheme ensures that the estimator follows the high-precision \texttt{BCNZ} solution at bright magnitudes, defaults to the more reliable broad-band solution at the faint end, and even slightly outperforms both around the transition region near $i_{\mathrm{AB}}=22.5$ for the PAUS data \citepalias{navarro-gironesPAUSurveyPhotometric2023}.

For the broad-band photo-$z$s from the template-fitting photo-$z$ algorithm \texttt{BPZ}, we use the $ugriz$ bands or the $ugriZYJHK_{\rm s}$ bands to mimic CFHTLenS or KiDS, respectively. 
For the data, we use the archival \texttt{BPZ} photo-$z$ catalogues from CFHTLenS \citep{hildebrandtCFHTLenSImprovingQuality2012} and KiDS \citep{wrightFifthDataRelease2024}. In both cases, \texttt{BPZ} was run using modified \citet{capak2004} templates. For the mocks, we rerun \texttt{BPZ} on the simulated photometry using the same modified template set. We adopt the KiDS prior configuration \citep{raichoorNextGenerationVirgo2014}, which differs slightly from the original CFHTLenS implementation. However, rerunning \texttt{BPZ} on the CFHTLenS data with this setup did not lead to significant changes in the resulting photo-$z$ statistics.

\section{\label{sc:Results}Mock photo-$z$ catalogues}
The simulated catalogues allow us to test the photo-$z$ algorithms in a fully controlled setting with knowledge of the true redshifts for all galaxies in the sample. 
In this way, it is also possible to isolate how survey properties, such as photometric depth, filter coverage, and noise levels, affect the performance of photometric redshifts in PAUS-like surveys. In this section, we test the realism of the simulated photo-$z$ catalogues and introduce a noise modification to properly mimic the observed properties.

\subsection{Comparing to data \label{sec:adding_noise}}
First, to confirm the realism of the mocks, we compare their photo-$z$ statistics with the data. For a fair comparison, we chose a subsample of simulated galaxies that is similar to the validation sample of the real data, which were the PAUS galaxies with spectroscopic redshifts. 
Such a subsample, dubbed `matched', was presented in Sect.\,\ref{sec:matched_sample}.

\begin{figure} [h]
    \centering
    \includegraphics[width=0.5\textwidth]{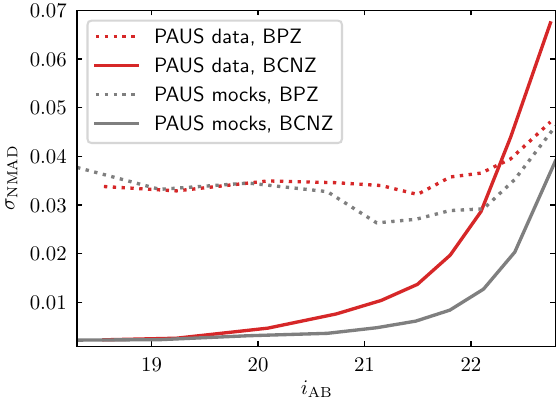}
    \caption{
    Photo-$z$ scatter as a function of $i_{\rm AB}$, binned in equal numbers of objects. We compare the data (red) and the matched mocks (grey), specifically the \texttt{BCNZ} photo-$z$s (solid lines) and the \texttt{BPZ} photo-$z$s (dotted lines), to analyse the differences and similarities in these statistics. The \texttt{BPZ} results are obtained using only the external broad-band photometry for galaxies located in the PAUS footprint, while \texttt{BCNZ} additionally uses the PAUS narrow-band data. Generally, there are clear deviations between the data and the mocks for \texttt{BCNZ}, indicating that we need to adjust the noise level of the narrow-band fluxes. 
    }
    \label{fig:photo-z_statistics_default_matched_scatter}
\end{figure}

Figure~\ref{fig:photo-z_statistics_default_matched_scatter} shows the photo-$z$ scatter as normalised median absolute deviation $\sigma_{\sfont{NMAD}}$ (Eq.\,\ref{eq:nmad}) as a function of the $i_{\rm AB}$ magnitude. We compare two algorithms: \texttt{BCNZ} photo-$z$s (40 narrow bands plus broad bands, solid lines), and \texttt{BPZ} photo-$z$s (broad bands only, dotted lines). Results are shown both for the PAUS data (red) and for the simulated matched mock catalogue (grey) at the baseline noise level, that is, with photometric uncertainties matching those expected for the PAUS survey under its standard observing conditions (see Fig.\,\ref{fig:flux_comp}). 

\begin{figure*}
    \centering
    \includegraphics[width=\textwidth]{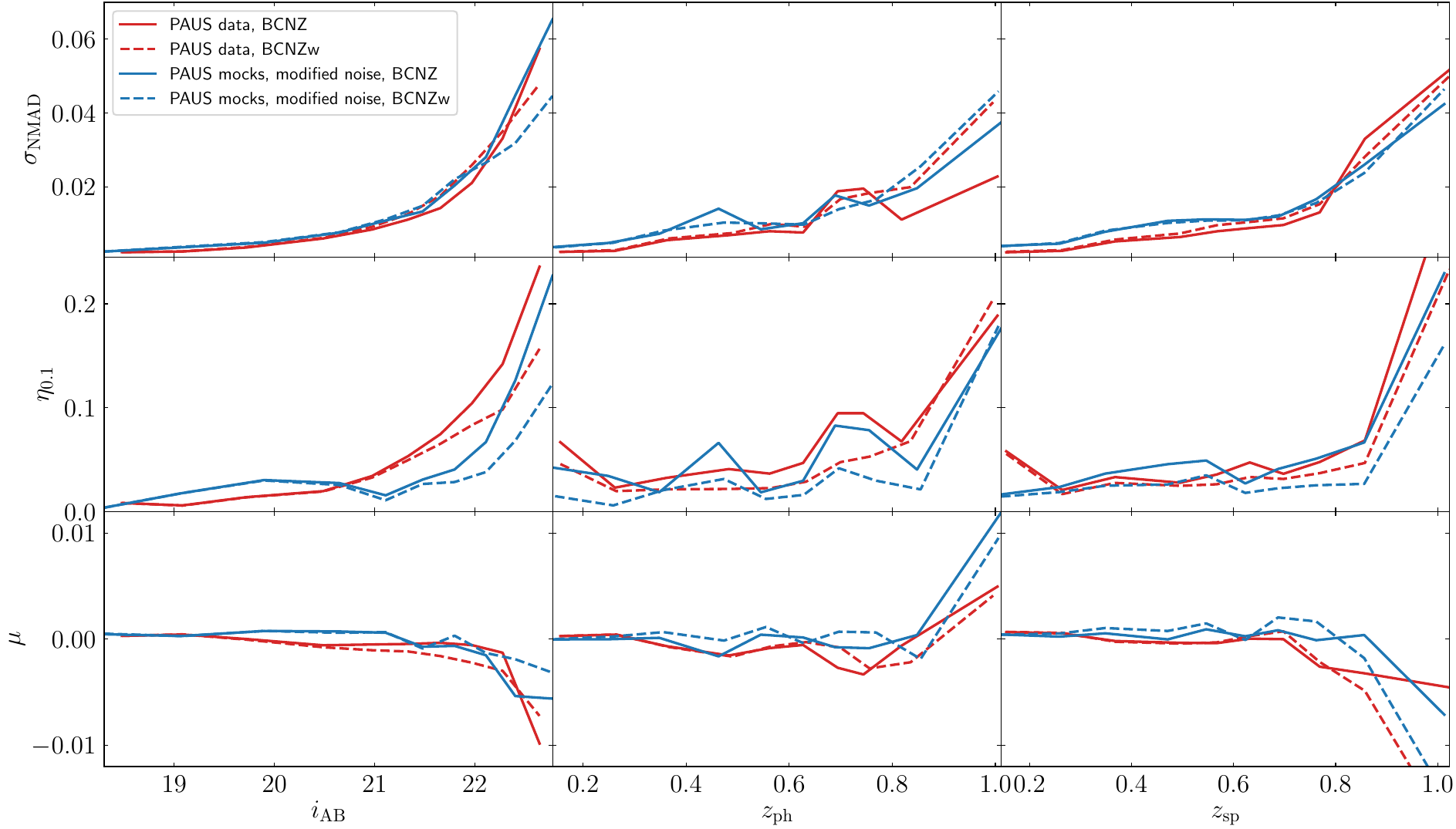}
    \caption{Photo-$z$ scatter (top), outlier fraction (middle), and photo-$z$ bias (bottom) as a function of $i_{\rm AB}$ (left), $z_{\rm ph}$ (middle), and $z_{\rm sp}$ (right), binned in equal numbers of objects. We compare the data (red) and the matched mocks with the added noise (blue, explained in Sect.\,\ref{sec:adding_noise}), specifically the \texttt{BCNZ} photo-$z$s (solid lines) and the weighted version, \texttt{BCNZ}w photo-$z$s (dashed lines). We find much better agreement with the adjusted noise level for the narrow-band fluxes.
    }
    \label{fig:photo-z_statistics_matched}
\end{figure*} 

Comparing the \texttt{BPZ} measurements of the mocks and the data in Fig.\,\ref{fig:photo-z_statistics_default_matched_scatter}, we see that they agree well, with differences smaller than 0.01 in most bins. In contrast, the \texttt{BCNZ} measurements from the mocks tend to underestimate the $\sigma_\sfont{NMAD}$ relative to the data, particularly at faint magnitudes ($i_{\rm AB}>22$), where the discrepancy grows to about 0.03. 

The fact that the \texttt{BCNZ} mocks give lower scatter than the data, while the broad-band based \texttt{BPZ} photo-$z$s show good agreement, suggests that the noise in the mock narrow bands might be underestimated (see also Appendix\,\ref{appendix:default_noise} for more details).
Additional sources of flux uncertainty, not captured by the current noise model (e.g., calibration errors, aperture effects, sky residuals, scatter light, or blending), might have affected the real observations. 

Another reason for the low photo-$z$ scatter could be the template sets that are used for \texttt{BCNZ} photo-$z$ estimation and the Flagship 2 mocks. Template-fitting photo-$z$ methods rely on a finite set of SED templates that represent only an approximate description of the full diversity of galaxy spectra. The Flagship~2 mocks are likewise based on galaxy SED libraries derived from COSMOS/Polletta and BC03 templates, which are also related to the template basis used by \texttt{BCNZ}. This partial consistency between the simulated galaxy population and the fitting templates can reduce the mismatch between observed and model fluxes, leading to improved photo-$z$ performance compared to real data.

However, the implementations are not identical. In Flagship~2, galaxy SEDs are generated from interpolated combinations of templates with extinction distributions calibrated to COSMOS2020 observations, while \texttt{BCNZ} uses a restricted set of template groups and attenuation configurations optimised for photo-$z$ estimation. The treatment of emission lines also differs significantly: Flagship~2 computes line fluxes from physically motivated relations involving star-formation rate, metallicity, extinction, and empirical calibrations, whereas \texttt{BCNZ} models emission lines through additional phenomenological templates with fixed relative line ratios.

To mitigate the discrepancy in the photo-$z$ statistics of the data and the mocks, we explore an empirical rescaling of the noise, effectively broadening the flux error distributions until the mock photo-$z$ statistics match the observed data more closely. This rescaling should not be interpreted as a correction to the photometric noise model alone, but rather as an effective proxy for several sources of uncertainty and systematic effects that are not fully captured in the simulations, such as incomplete SED diversity in the template set, residual photometric calibration uncertainties, blending and aperture effects, zero-point variations, or contamination from stars and AGN. We note that increasing the random noise does not fully reproduce the impact of such systematics, which can affect the photo-$z$ distributions and outlier fractions differently.
One particular statistic is the scatter as a function of the $i_{\rm AB}$ magnitude (see Fig.\,\ref{fig:photo-z_statistics_default_matched_scatter}), which can be brought to agreement by increasing the noise of the mocks by a factor of 2 (or equivalently decreasing the input 5\,$\sigma$ limiting magnitude by $\Delta m_{\rm lim} = 0.75$).
The resulting statistics, shown in Fig.\,\ref{fig:photo-z_statistics_matched}, confirm that the photo-$z$ performance shows much better agreement with the data, especially for $z_{\rm \texttt{BCNZ}}$ (solid lines). 

\begin{figure*}
    \centering
    \includegraphics[width=\textwidth]{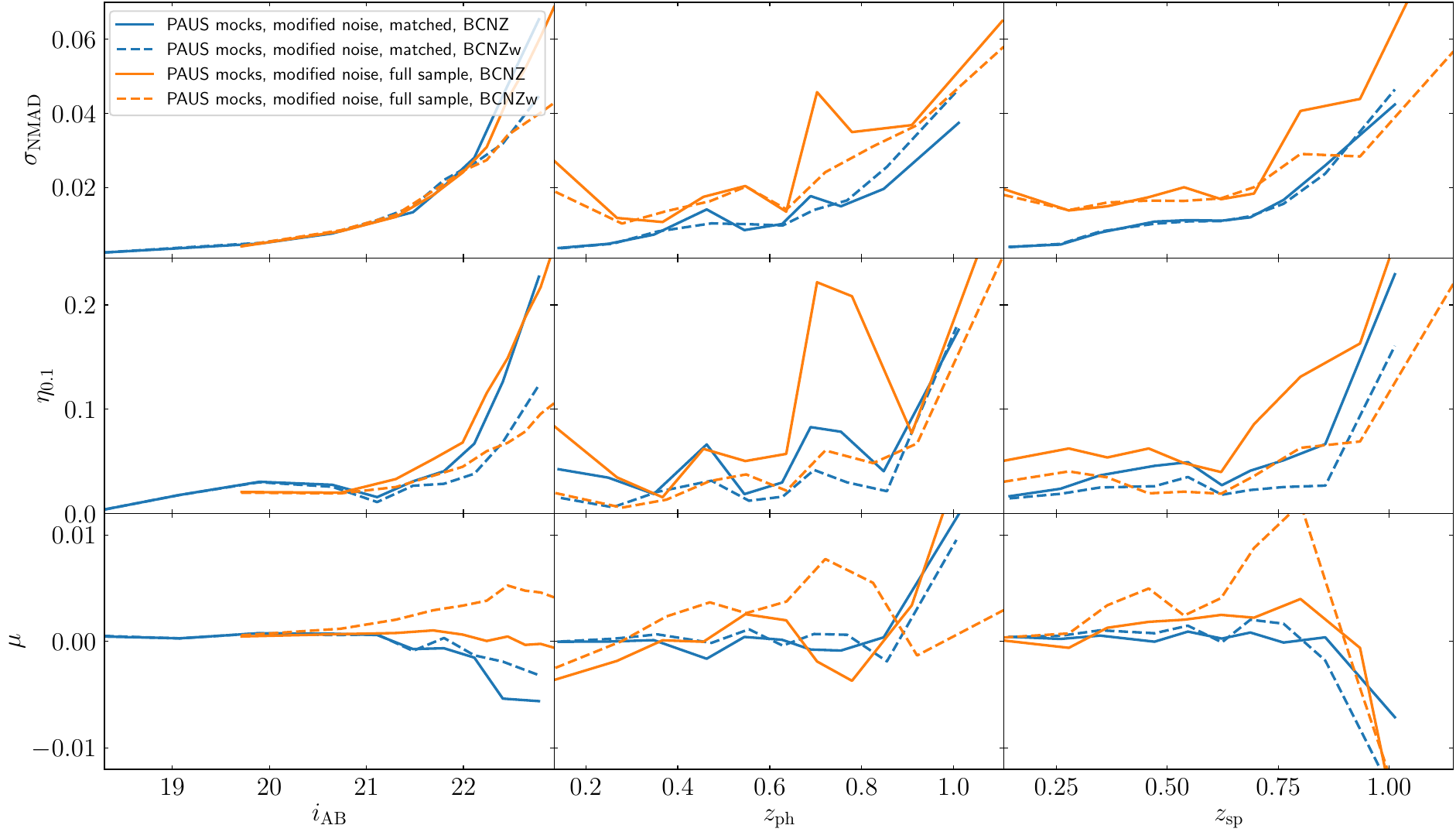}
    \caption{Same as Fig.\,\ref{fig:photo-z_statistics_matched}, but now we compare the matched mocks and the full sample, both with the added noise, specifically the \texttt{BCNZ} photo-$z$s (solid lines) and the weighted version, \texttt{BCNZ}w photo-$z$s (dashed lines).
    }
    \label{fig:photo-z_statistics_full}
\end{figure*} 

\subsection{Weighted photometric redshifts}

To assess whether the simulations reproduce the behaviour of more advanced photo-$z$ estimators, we analyse a weighted combination of \texttt{BCNZ} and \texttt{BPZ}, referred to as \texttt{BCNZ}w and introduced in Sect.~\ref{sec:weighted_photo-z}. 
Figure\,\ref{fig:photo-z_statistics_matched} shows the photo-$z$ performance metrics, which were introduced in Sect.\,\ref{sec:statistics}. From top to bottom, the photo-$z$ scatter as normalised median absolute deviation $\sigma_{\sfont{NMAD}}$ (Eq.\,\ref{eq:nmad}), the outlier fraction $\eta_{\sfont{0.1}}$ (Eq.\,\ref{eq:outlier}), and photo-$z$ bias $\mu$ (Eq.\,\ref{eq:mu}) are shown. To compute these metrics, we show them as a function of $i_{\rm AB}$ magnitude, the $z_{\rm ph}$ as the photometric redshift measured from the synthetic catalogue and the observations, and compare them with $z_{\rm true}$, defined as the true redshift from the mocks and the spectroscopic redshift $z_{\rm sp}$ for the data, respectively. We compare the data (red) with the matched mocks (blue). Solid lines correspond to the \texttt{BCNZ} photo-$z$ estimates, while dashed lines show the weighted \texttt{BCNZ}w photo-$z$s.  

Overall, the mocks reproduce the observed trends well, improving substantially over the results shown in Fig.\,\ref{fig:photo-z_statistics_default_matched_scatter} (and Fig.\,\ref{fig:photo-z_statistics_default_matched}). 
For the photo-$z$ scatter in the \texttt{BCNZ} case, the values increase with magnitude from around $5 \times 10^{-4}$ at $i_{\rm AB} =19$ to 0.08 at $i_{\rm AB} = 23$ in both data and mocks. The weighted \texttt{BCNZ}w photo-$z$s show slightly better performance, with the mocks having marginally lower values for the photo-$z$ scatter than the data. As a function of $z_{\rm true}$ and $z_{\rm ph}$, the $\sigma_\sfont{NMAD}$ value typically lies in the range $0.005$--$0.05$, with mocks and data following each other closely for both \texttt{BCNZ} and \texttt{BCNZ}w. 

The outlier fraction increases with magnitude, reaching 25\% in the faintest bin. The mocks reproduce this trend well, with the weighted case (dashed lines) yielding consistently lower outlier fractions than the unweighted \texttt{BCNZ} estimates. Across redshift, the outlier fraction remains below $0.1$ for most bins, with the mocks systematically performing slightly better than the data. 
Outliers represent one of the main challenges for photo-$z$ applications in cosmology, since they can bias clustering and lensing analyses. In our mocks, we observe that the outlier population is dominated by galaxies at the faint end ($i_{\rm AB} \gtrsim 22$) where photometric noise significantly reduces the ability to localise spectral features. Another contributing factor is galaxies with extreme and complex SEDs, for which the template library provides potentially imperfect matches.

The photo-$z$ bias remains close to zero for both \texttt{BCNZ} and \texttt{BCNZ}w, typically within $|\mu| < 0.005$ across magnitude and redshift. The bias only increases at higher redshifts ($z_{\rm sp}>0.8$) or photo-$z$s ($z_{\rm ph}>0.8$), reaching values of $|\mu| > 0.01$. 
More details on the differences between \texttt{BCNZ} and \texttt{BCNZ}w are also presented in Appendix\,\ref{appendix:complete_samples}. 
Taken together, these results demonstrate that the mock catalogues successfully reproduce the previously established behaviour of the weighted \texttt{BCNZ}w estimator and provide a reliable framework for testing and validating hybrid photo-$z$ approaches that combine narrow-band and broad-band information. 

\subsection{Complete samples}
Finally, we note that the selection of galaxies for the validation samples can influence the apparent photo-$z$ statistics. The spectroscopic validation sets are incomplete at faint magnitudes ($i_{\rm AB} > 22$) and high redshifts ($z_{\rm sp} > 1$) for the real data, where most of the galaxies of the complete sample would be. One of the main reasons why simulations are so instructive is the advantage of complete knowledge of all galaxies down to the flux limit of the PAUS data. By comparing the photo-$z$ performance of the complete simulated sample versus the spectroscopic-matched subsample, we can quantify the impact of these selection effects. 

Figure\,\ref{fig:photo-z_statistics_full} shows the statistics for the simulated PAUS wide-field sample, which are only slightly degraded compared to the statistics of the matched sample. Some systematic differences are visible for \texttt{BCNZ}. One dominant effect is the accumulation of outliers for $z_{\rm ph}=0.7$--0.8, which are likely caused by the effect of `redshift focussing' (see also Figs.\,\ref{fig:scatter_plot_matched} and \ref{fig:scatter_plot_all}). The accumulation point is also visible in the real PAUS data and was discussed by \citetalias{navarro-gironesPAUSurveyPhotometric2023}. It occurs when galaxies with low signal-to-noise narrow-band photometry have broad and weakly constraining likelihood functions, causing the inferred redshifts to cluster around regions favoured by the priors or template degeneracies. In the case of \texttt{BCNZ}, the accumulation point corresponds approximately to the centre of the effective redshift prior for the CFHTLenS and KiDS broad-band configuration (more details in Appendix\,\ref{appendix:complete_samples}).  
Emission-line modelling may also contribute to this behaviour, because mismatches in the treatment of emission lines can enhance template degeneracies and are known to produce accumulation points in photo-$z$ estimates. We have not isolated this effect in the present analysis, but it may act in combination with the prior-driven redshift focussing.

The accumulation effect is reduced for the weighted estimator \texttt{BCNZ}w.
Thus, the improvement of the statistics for the weighted photometric redshift estimates is particularly prominent for the complete sample, reducing the overall outlier rates by nearly 5\%. We learn that the reported scatter and outlier rates in real PAUS analyses with the incomplete spec-$z$ validation sample are fairly realistic, especially when using the weighted photo-$z$ estimator \texttt{BCNZ}w. Nevertheless, caution is needed when extrapolating photo-$z$ statistics far beyond the range directly constrained by spectroscopy. 

To summarise the findings, Table\,\ref{tab:photoz_summary_matched} presents the mean statistics for all the different scenarios discussed in the previous subsections. The statistics derived for the full sample are generally slightly worse because we average over the full magnitude range up to $i_{\rm AB}=23$, while the matched sample is brighter on average. For the full sample, we also see the effect of the weighted estimate \texttt{BCNZ}w, which reduces the photo-$z$ scatter and the outlier fraction compared to the unweighted \texttt{BCNZ} photo-$z$s.

\begin{table*}[htbp!]
\centering
\caption{Summary of photo-$z$ statistics for different survey configurations. We report the mean $i_{\rm AB}$ of the sample, the photo-$z$ scatter $\sigma_\sfont{NMAD}$, the fraction of catastrophic outliers $\eta_{\sfont{0.1}}$, and the bias $\mu$. These are the mean values for a mock sample with a depth of $i_{\rm AB}<23$, directly comparable to the statistics of the real PAUS data.}
\smallskip
\label{tab:photoz_summary_matched}
\smallskip
\begin{tabular}{lllcccc}
\hline\hline
\noalign{\vskip 2pt}
Configuration &  & $z_{\rm ph}$\,estimation & $\langle i_{\rm AB}\rangle$ & $\sigma_\sfont{NMAD}$ & $\eta_{\sfont{0.1}}$ & $\mu$ \\
\hline
  \noalign{\vskip 3pt}
PAUS data &  & \texttt{BCNZ} & 20.86 & $\phantom{0} 7.20 \times 10^{-3}$ & \phantom{0} 6.9\% & $  - 2.0 \times 10^{-4}$ \\
PAUS data &  & \texttt{BCNZ}w & 20.86 & $\phantom{0}9.17 \times 10^{-3}$ & \phantom{0} 5.3\% & $ - 3.6 \times 10^{-4}$ \\
PAUS mocks, matched &  & \texttt{BCNZ} & 20.96 & $\phantom{0} 5.43 \times 10^{-3}$ & \phantom{0} 3.4\% & $\phantom{0} 1.2 \times 10^{-4}$ \\
PAUS mocks, matched & mod. noise & \texttt{BCNZ} & 20.96 & $ 10.56 \times 10^{-3}$ & \phantom{0} 5.9\% & $ \phantom{0} 1.9 \times 10^{-4}$ \\
PAUS mocks, matched & mod. noise & \texttt{BCNZ}w & 20.96 & $11.36 \times 10^{-3}$ & \phantom{0} 3.7\% & $\phantom{0} 3.7 \times 10^{-4}$ \\
PAUS mocks, full sample & mod. noise & \texttt{BCNZ} & 21.84 & $32.04 \times 10^{-3}$ & \phantom{\,} 14.4\% & $ 11.4 \times 10^{-4}$ \\
PAUS mocks, full sample & mod. noise & \texttt{BCNZ}w & 21.84 & $28.47 \times 10^{-3}$ & \phantom{0} 9.2\% & $ 29.7 \times 10^{-4}$ \\
\hline
\end{tabular}
\end{table*}

\section{\label{sec:deeper_paus} Exploring the impact of PAUS depth and including external data}

So far, our analysis has focused on mocks designed to reproduce the existing PAUS wide survey.  
As an extension, it is instructive to explore how photo-$z$ performance would change if the PAUS narrow-band observations were taken to greater depth.  
Additionally, as was already presented in the study by \citet{alarconPAUSurveyImproved2021}, the broad-band photometry for the computation of the photo-$z$s can be expanded, also resulting in greater depth or a longer wavelength baseline than the current configuration.
Although a deeper PAUS sample and/or combinations with deeper broad-band photometry are not part of the current survey strategy, it highlights the role of depth in narrow-band surveys and illustrates the potential of future facilities that may adopt a similar approach. 

\subsection{Simulating PAUS deep}
\label{sec:deep_paus}
With the simulations, we can explore how adjusting survey parameters affects the quality of photo-$z$ estimates from narrow-band photometry.  
At the current PAUS magnitude limits, it is feasible to obtain spectroscopic redshifts for an extremely large number of galaxies with existing instruments, as demonstrated by surveys such as the Dark Energy Spectroscopic Instrument \citep[DESI,][]{martiniOverviewDarkEnergy2018} and the 4-metre Multi-Object Spectroscopic Telescope \citep[4MOST,][]{jong4MOSTProjectOverview2019}.  
At fainter magnitudes, however, spectroscopic redshift measurements become prohibitively expensive: longer exposures are required, and the decreasing photon counts make reliable measurements increasingly inefficient \citep{newmanSpectroscopicNeedsImaging2015}.  
In this regime, photometric narrow-band surveys can provide an attractive alternative, potentially offering quasi-spectroscopic information for large samples of galaxies. In particular, the number density can be increased through the use of photometric measurements compared to spectroscopy, which offer opportunities for interesting science cases such as cross-correlation measurements. 
Here, we investigate a forecast scenario for a deeper version of PAUS. 

To evaluate the potential gain in photometric depth from upgraded observing facilities, we consider a transition from the 4.2-metre WHT to an 8-metre-class telescope, coupled with a doubling of exposure time. The typical PAUS observations consist of three exposures of 2–3 minutes per pointing \citep{serranoPAUSurveyNarrowband2023}, with eight narrow-band filters observed simultaneously in each exposure through the five PAUCam filter trays. 
The corresponding improvement in limiting magnitude can be estimated as
\begin{equation}
\Delta m_{\rm lim} \simeq 2.5 \logten\! \sqrt{\frac{D_{\rm new}^2 \, t_{\rm new}}{D_{\rm PAUS}^2 \, t_{\rm PAUS}}}  \approx 1.08 \;,
\end{equation}
with $D$ being the diameter and $t$ the exposure time. 

We further assume an improvement in typical image quality, from $1.\!^{\prime\prime}0$–$1.\!^{\prime\prime}1 \pm 0.\!^{\prime\prime}2$ at the WHT site to $0.\!^{\prime\prime}8 \pm 0.\!^{\prime\prime}1$ as achieved at premier observing sites such as Maunakea, Hawaii, or the Atacama Desert, Chile.  
The smaller PSF concentrates flux into fewer pixels, increasing the S/N at a given magnitude.  
Together, these improvements substantially extend the effective depth and sensitivity of the survey in a potential future observing campaign (see Fig.\,\ref{fig:mag_limits_psf_data}).   

For a more realistic test, we apply the improvement in limiting magnitude to the PAUS sample with the modified noise modelling.
The deeper version of PAUS, labelled `PAUS deep', includes fainter fluxes and reduces flux uncertainties due to improved image quality and a higher limiting magnitude.  
As a result, the S/N distribution shifts upward for all narrow bands (see Fig.\,\ref{fig:flux_deep}).  
This experiment highlights the potential of future narrow-band surveys on larger aperture telescopes, where the combined effects of greater depth and improved seeing could greatly enhance the reach of high-precision photo-$z$ techniques at fainter magnitudes.

\begin{figure*}[h]
    \centering
    \includegraphics[width=\linewidth]{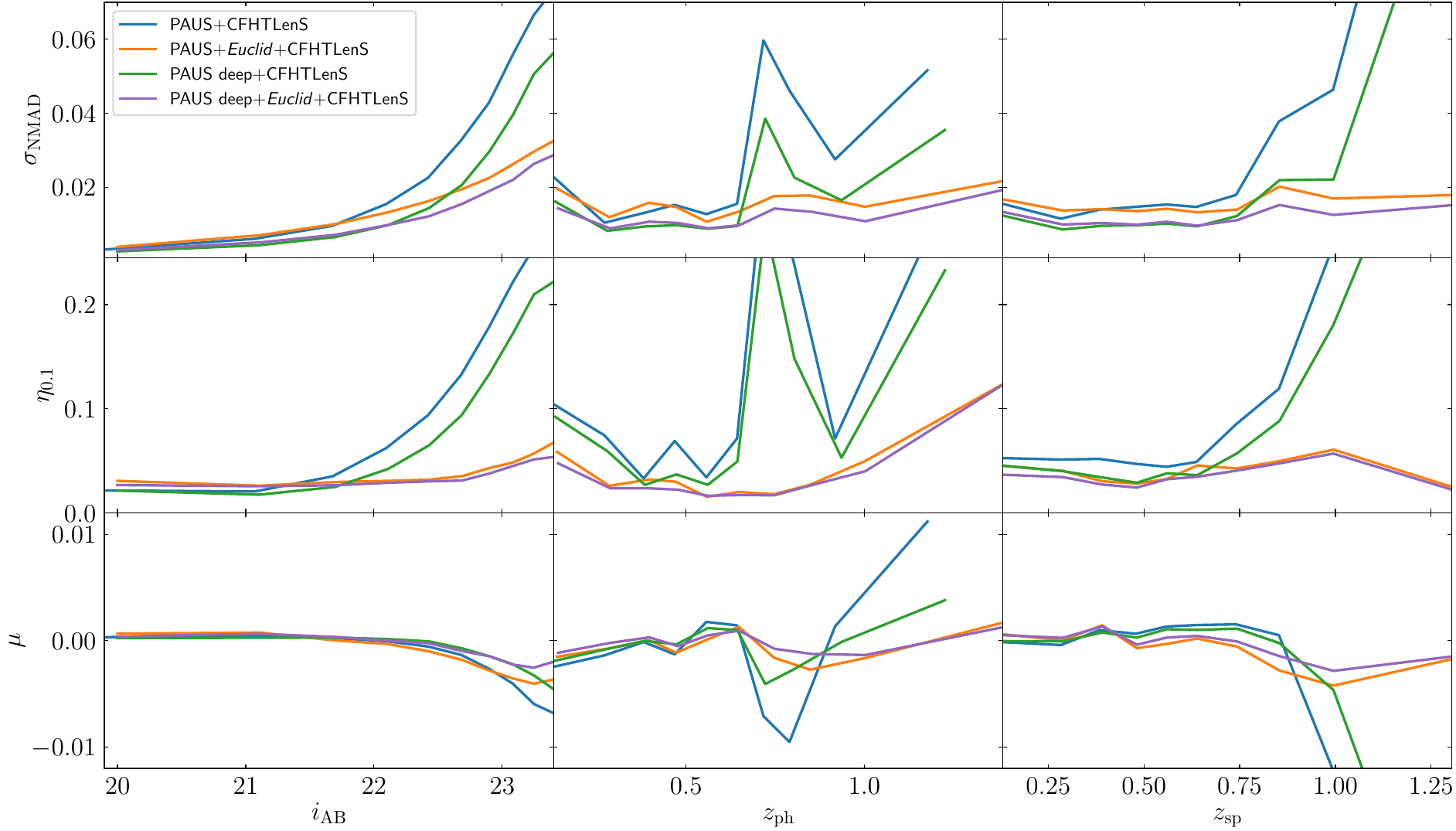}
    \caption{Same as Fig.\,\ref{fig:photo-z_statistics_matched}, but now we compare the redshift statistics for hypothetical configurations, also including \Euclid broad-band photometry. All results use \texttt{BCNZ} photo-$z$s. Shown are complete PAUS mocks with CFHTLenS-like photometry, with and without \Euclid-like photometry, and the same configurations for the deeper simulated version of PAUS. The PAUS deep cases highlight the potential gains of future narrow-band surveys with greater depth.}
    \label{fig:photoz_statistics_euclid}
\end{figure*}

\subsection{Forecasts including CFHTLenS photometry}
For a more insightful comparison, we now analyse all galaxies in the mock sample down to $i_{\rm AB}<23.5$, rather than restricting it to the selection of PAUS objects with spectroscopic redshifts. For the broad-band photometry, we use the simulated CFHTLenS photometry ($ugriz$) instead of a combination of CFHTLenS and KiDS broad-band photometry. 

The resulting photo-$z$ statistics, shown in Fig.\,\ref{fig:photoz_statistics_euclid}, therefore include a much larger fraction of faint sources. This analysis is restricted to $i_{\rm AB}<23.5$ for reasons of computational efficiency in the photo-$z$ estimation. The number of mock galaxies rises steeply at fainter magnitudes; however, this magnitude limit still provides an interesting range of S/N to assess depth-dependent trends. We only use the \texttt{BCNZ} photo-$z$ estimates given that, in this test, we are mostly interested in the effect of the depth of the narrow-band photometry. 

The sample `PAUS+CFHTLenS' (see Fig.\,\ref{fig:photoz_statistics_euclid}) shows degraded performance compared to the standard PAUS configuration in Fig.\,\ref{fig:photo-z_statistics_matched}, through the statistics plotted, as a function of $z_{\rm ph}$ and $z_{\rm sp}$. This is expected since we now include more faint objects.  
When we consider the deeper PAUS configuration combined with CFHTLenS $ugriz$ photometry, `PAUS deep+CFHTLenS', we find that the increased narrow-band depth reduces photo-$z$ scatter and the fraction of catastrophic outliers in nearly all cases, compared to PAUS+CFHTLenS. 

This test illustrates that survey depth is critical and that the highly resolved information encoded in narrow bands can be fully leveraged with \texttt{BCNZ} when the S/N is sufficiently high. Figure\,26 of \citet{serranoPAUSurveyNarrowband2023} already showed this behaviour, where they found that the photo-$z$ scatter decreases by more than a factor of 1.5 when doubling the narrow-band S/N. The photo-$z$s for emission-line galaxies benefit strongly from deeper data, since their spectral features can be precisely localised in the narrow bands, while quiescent galaxies also improve due to higher continuum S/N. 

\subsection{Forecasts including \Euclid + CFHTLenS photometry}
\label{sec:paus_euclid_cfht}

Maximising the scientific potential of \Euclid requires combining its photometric measurements with external data sets. While photometry from UNIONS and LSST are the default complements to the final data release of \Euclid \citep{Scaramella-EP1}, and are used for most forecasts \citep[e.g.,][]{EP-Humphrey2, Q1-TP005}, 
we specifically evaluate the benefits of adding PAUS narrow-band and CFHTLenS broad-band information for the estimation of photometric redshifts. Although CFHTLenS spans a much smaller sky area, its depth, which is comparable to UNIONS and LSST, makes it a compelling data set to analyse. The PAUS fields W3, W1, and G09 all lie within the \Euclid survey footprint, making such combinations observationally relevant. In particular, the W3 field is already covered by the first \Euclid data release (DR1).

With the simulations, we can assess the impact of combining \Euclid with PAUS data and with a deeper version of that survey, as explained in Sect.\,\ref{sec:deep_paus}, where the higher S/N would enable the quasi-spectroscopic information to be fully exploited. Figure\,\ref{fig:mag_limits_psf_data} shows the expected limiting magnitudes and seeing for \Euclid \citep{Scaramella-EP1}. 
For the forecasts presented below, we utilise the noise realisations of the \Euclid photometry that are provided by Flagship 2. 
Although these noisy fluxes were created before the first observations of \Euclid, they provide a realistic approximation of its performance.  

We combine PAUS at its standard depth with simulated CFHTLenS and \Euclid broad-band photometry ($\YE$, $\JE$, and $\HE$). In a previous study, \citet{Cabayol23} showed that adding PAUS narrow-band photometry to \Euclid broad-band data within a multi-task learning framework improves \Euclid-based photometric redshifts, yielding a 13\% gain in precision and a 40\% reduction in outlier rate for galaxies with $i_{\rm AB}<23$. In contrast, in our mock-based analysis, we examine the PAUS photo-$z$ performance under varying depth levels and including \Euclid photometry, while using template-based photo-$z$ estimation for $i_{\rm AB}<23.5$.
As shown in Fig.\,\ref{fig:photoz_statistics_euclid}, the addition of \Euclid reduces the scatter $\sigma_\sfont{NMAD}$ by at least 50\% relative to PAUS+CFHTLenS at $i_{\rm AB}>22$ and high redshifts, caused by the extension of the continuum coverage into the near-infrared.  
It also significantly lowers the outlier fraction, particularly mitigating the effects of redshift focussing around $z_{\rm ph}=0.7$--0.8.  

Finally, we examine the combination of the deeper PAUS configuration with both \Euclid and CFHTLenS (`PAUS deep+$\Euclid+$CFHTLenS').  
This represents an idealised scenario, providing an upper limit on performance achievable when narrow-band depth and extended wavelength coverage are available simultaneously. 
The results indicate photo-$z$ precision comparable to the `PAUS+CFHTLenS+\Euclid' case, but with a further reduction in photo-$z$ scatter.  
This shows that the addition of \Euclid photometry further improves the quality of the photo-$z$s in either case of the PAUS photometry. 
In this test, we only derive the statistics for $i_{\rm AB}<23.5$, but it suggests that for an even fainter sample, including the \Euclid photometry would also be beneficial. 

\begin{table*}
\centering
\caption{Summary of photo-$z$ statistics for different combinations of broad- and narrow-band photometry. We report the mean $i_{\rm AB}$ of the sample, the photo-$z$ scatter $\sigma_\sfont{NMAD}$, the fraction of catastrophic outliers $\eta_{\sfont{0.1}}$, and the bias $\mu$. These are the mean values for a mock sample with a depth of $i_{\rm AB}<23.5$, all based only on simulations. }
\smallskip
\label{tab:photoz_summary_euclid}
\smallskip
\begin{tabular}{lcccc}
\hline\hline
\noalign{\vskip 2pt}
Configuration & $\langle i_{\rm AB}\rangle$ & $\sigma_\sfont{NMAD}$ & $\eta_{\sfont{0.1}}$ & $\mu$ \\
\hline
  \noalign{\vskip 3pt}
PAUS+CFHTLenS & 22.25 & $ 34.66 \times 10^{-3}$ & 17.4\% & $ -10.1\times 10^{-4}$ \\
PAUS+\Euclid+CFHTLenS & 22.25 & $ 20.32 \times 10^{-3}$ & \phantom{0}4.6\% & $ - 14.5\times 10^{-4}$ \\
PAUS deep+CFHTLenS & 22.25 & $ 13.61 \times 10^{-3}$ & 10.0\% & $\phantom{0}-  2.3\times 10^{-4}$ \\
PAUS deep+\Euclid+CFHTLenS & 22.25 & $ 11.69 \times 10^{-3}$ & \phantom{0}3.6\% & $ \phantom{0}-2.1\times 10^{-4}$ \\
\hline
\end{tabular}
\end{table*}

\subsection{Summary of PAUS deep forecasts}
These forecast experiments underline the importance of both survey depth and complementarity with broad-band data.  
They show that narrow-band imaging at fainter limits can approach quasi-spectroscopic photo-$z$ precision for large galaxy samples, and that adding near-infrared photometry greatly improves performance across all ranges and reduces degeneracies and systematic errors. One very prominent feature in Fig.\,\ref{fig:photoz_statistics_euclid} is the outlier rate, scatter and bias at $z_{\rm ph}=0.7$--0.8. With only PAUS+CFHTLenS, we can observe the effect of photo-$z$ focussing, when the noisy flux measurements result in prior-dominated photometric redshift estimation. With increased depth in the PAUS fluxes or, even more effectively, when adding the \Euclid photometry, this focussing effect gets alleviated, and in particular the outlier rate in this parameter space reaches levels of a few per cent (instead of above 25\%). This is also demonstrated in the photo-$z$ scatter plots in Appendix\,\ref{appendix:forecasts}.

Although the current PAUS wide survey is shallower than the nominal depth of the \Euclid weak-lensing sample, PAUS-like narrow-band observations can still provide important complementary information for \Euclid science. In particular, the quasi-spectroscopic redshift precision achievable with narrow-band photometry is valuable for the calibration and validation of broad-band photo-$z$ estimates, the identification and characterisation of catastrophic outliers, and improved SED reconstruction in deep calibration fields. Such data can serve as high-quality training or reference samples for empirical and hybrid photo-$z$ methods, complementing spectroscopic calibration efforts. Furthermore, joint narrow- and broad-band observations are highly relevant for legacy science applications, including galaxy evolution studies.
While the current PAUS data already demonstrate the viability of the narrow-band approach, these simulations highlight the opportunities for future surveys that expand upon the PAUS concept. 
To complement the qualitative comparisons, Table\,\ref{tab:photoz_summary_euclid} summarises key metrics for the different survey configurations.  
The results highlight the progressive gains achieved by adding depth and complementary data. 
This extends reliable photo-$z$ performance to fainter magnitudes, substantially increasing the usable galaxy sample for many interesting science cases. 
For cosmology, reducing scatter and outlier rates directly increases the effective number density of galaxies available for large-scale structure and weak lensing analyses, strengthening statistical power.  
For galaxy evolution, deeper narrow-band data extend emission-line detection to fainter systems, enabling studies of galaxies at higher redshifts and lower stellar masses than currently accessible. 

\section{\label{sc:Conclusion}Conclusions}

The accurate measurement of galaxy redshifts remains a central challenge for cosmology and galaxy evolution.  
While spectroscopy provides the benchmark, it becomes inefficient at faint magnitudes, motivating the development of alternative techniques.
PAUS, with its unique set of narrow-band filters, offers a promising approach towards quasi-spectroscopic photo-$z$s.  

We have constructed and analysed mock catalogues designed to reproduce the photometric and redshift properties of PAUS-like narrow-band surveys.  
These simulations provide a fully controlled environment to test photo-$z$ algorithms, allowing us to disentangle the effects of photometric depth, filter configuration, and noise from the intrinsic properties of the galaxy population.  
By comparing the mock catalogues directly to PAUS data \citepalias{navarro-gironesPAUSurveyPhotometric2023}, we have evaluated both the realism of the simulations and the performance of different photo-$z$ estimators.

Introducing an empirical increase in the noise level by a factor of 2 brings the mock photo-$z$ statistics into closer agreement with the data.  
This adjustment demonstrates the sensitivity of photo-$z$ performance to changes in photometric precision.

We further investigated a weighted photo-$z$ estimator (\texttt{BCNZ}w), which combines the strengths of two photo-$z$ algorithms: \texttt{BCNZ} and \texttt{BPZ}.  
This hybrid approach improves both the stability of the photo-$z$ bias and the control of catastrophic outliers, particularly at faint magnitudes and higher redshifts, where single-method estimates become unreliable.  
\texttt{BCNZ}w therefore provides the most balanced performance across the full dynamic range of the survey, again demonstrating the advantage of combining narrow-band information with broad-band colour priors like in \citetalias{navarro-gironesPAUSurveyPhotometric2023}.

Analyses of the full simulated sample, free from spectroscopic selection biases, can test if the spec-$z$ validation sample potentially underestimates the true scatter and outlier fractions, especially for the faintest galaxies, due to the limited completeness of the spectroscopic sample. In this study, we found that the photo-$z$s estimated with \texttt{BCNZ}w better represent the full PAUS sample than those estimated with the unweighted estimator \texttt{BCNZ}. 
The complete mocks still offer an essential framework for calibrating and interpreting photo-$z$ statistics beyond the range directly constrained by spectroscopic data. 

Our forecasts of a deeper version of PAUS show that depth is the decisive factor in narrow-band surveys.  
A hypothetical `PAUS deep' scenario, representing observations on an 8-metre-class telescope with longer exposures and improved seeing, extends the limiting brightness by over one magnitude compared to the current PAUS setup.  
This gain translates into a reduction in photo-$z$ scatter, a marked decrease in catastrophic outlier rates, and reliable redshift estimates for a larger fraction of faint galaxies.  

Additional forecasts, including simulated photometry from \Euclid, showcase the crucial contribution of deep near-infrared measurements, which lead to improved photo-$z$ accuracy. This would already be apparent with the current PAUS data. The \Euclid Data Release 1 (DR1) does not overlap with the PAUS fields, but future data releases will enable the combined measurement of improved photo-$z$s.

These results highlight the complementarity between ground-based narrow-band and space-based near-infrared surveys: together, they can deliver both the spectral resolution and wavelength coverage required for precise and unbiased photometric redshifts across a wide redshift range. The present analysis is limited to $i_{\rm AB} < 23.5$ primarily for computational efficiency, but future studies that extend to fainter limits could further quantify the gains expected from a deeper narrow-band survey. Such an extension would be especially valuable for assessing the faint-end performance of photo-$z$ estimators in the context of next-generation facilities.

Although the present analysis is limited to $i_{\rm AB}<23.5$ and therefore does not yet probe the full depth of the \Euclid weak-lensing sample, the results illustrate several ways in which PAUS-like narrow-band data could support future \Euclid science analyses. High-precision narrow-band photo-$z$s can provide valuable calibration and validation samples for broad-band photo-$z$ methods, improve the reconstruction of galaxy SEDs, and help identify and characterise catastrophic outliers and redshift degeneracies. In addition, the quasi-spectroscopic redshift precision achievable with narrow bands offers opportunities for legacy science applications, including studies of galaxy evolution, emission-line populations, and clustering measurements in deep calibration fields. Future deeper narrow-band surveys overlapping with \Euclid could therefore complement the mission by providing high-quality redshift information in selected fields where spectroscopy is incomplete or observationally expensive.

At the same time, the forecasts are subject to important limitations.  
They are based on idealised simulations that do not yet account for systematic effects such as photometric calibration errors, filter transmission uncertainties, aperture effects, PSF variations, scattered light, or stellar contamination.  
In addition, the `PAUS deep' scenario assumes a gain in depth without addressing the inevitable trade-off with survey area under realistic observing constraints.  
As such, our results should be interpreted as idealised upper limits on performance rather than as predictions for a specific survey design.  

To summarise, our results demonstrate that the mock catalogues based on Flagship 2 are realistic and robust tools for testing photo-$z$ algorithms in PAUS-like surveys.  
They reproduce the main features of the observed photo-$z$ performance and enable systematic exploration of observational and algorithmic uncertainties.  
The hybrid \texttt{BCNZ}w estimator emerges as the most reliable configuration for achieving stable, unbiased photometric redshifts across a wide range of magnitude and redshift, underscoring the power of combined narrow- and broad-band photometry for future large-scale structure and weak-lensing analyses.

In conclusion, our study shows that extending the PAUS concept to greater depth could yield improvements in photometric redshift performance.  
By delivering quasi-spectroscopic redshifts for faint galaxies in selected deep fields, a deep narrow-band survey could complement surveys like \Euclid through improved photo-$z$ calibration, better control of catastrophic outliers, and enhanced SED reconstruction, while also opening new opportunities for galaxy evolution and cosmology studies.  
These forecasts demonstrate the potential of this approach to bridge the gap between broad-band photometry and spectroscopy.

%
%

\begin{acknowledgements}
HH is supported by a DFG Heisenberg grant (Hi 1495/5-1), the DFG Collaborative Research Center SFB1491, an ERC Consolidator Grant (No. 770935), and the DLR project 50QE2305.
DNG acknowledges support from the European Research Council (ERC) under the European Union’s Horizon 2020 research and innovation program with Grant agreement No. 101053992.
ME acknowledges support from the Spanish Ministry of Science and Innovation through the project PID2023-152069NA-I00.
JC acknowledges support from the Spanish Plan Nacional project PID2024-159420NB-C44.
FJC acknowledges support from MICIU/AEI/10.13039/501100011033 projects PID2022-141079NB-C31 and PCI2023-146001-2.
CP acknowledges support from the Spanish Plan Nacional project PID2022-141079NB-C32. 
PR acknowledges support from the Spanish Ministerio de Ciencia, Innovación y Universidades, through projects PID2022-138896NB; and the programme Unidad de Excelencia María de Maeztu, project CEX2020-001058-M.
The PAU Survey is partially supported by MINECO under grants CSD2007-00060, AYA2015-71825, ESP2017-89838, PGC2018-094773, PGC2018-102021, PID2019-111317GB, SEV-2016-0588, SEV-2016-0597, MDM-2015-0509 and Juan de la Cierva fellowship and LACEGAL and EWC Marie Sklodowska-Curie grant No 101086388 and no.776247 with ERDF funds from the EU Horizon 2020 Programme, some of which include ERDF funds from the European Union. IEEC and IFAE are partially funded by the CERCA and Beatriu de Pinos program of the Generalitat de Catalunya. Funding for PAUS has also been provided by Durham University (via the ERC StG DEGAS-259586), ETH Zurich, Leiden University (via ERC StG ADULT-279396 and Netherlands Organisation for Scientific Research (NWO) Vici grant 639.043.512), University College London and from the European Union's Horizon 2020 research and innovation programme under the grant agreement No 776247 EWC. The PAU data center is hosted by the Port d'Informaci\'o Cient\'ifica (PIC), maintained through a collaboration of CIEMAT and IFAE, with additional support from Universitat Aut\`onoma de Barcelona and ERDF. We acknowledge the PIC services department team for their support and fruitful discussions. 
\AckEC  
\AckCosmoHub
\end{acknowledgements}


\bibliography{Euclid,MyLibrary}

%

\begin{appendix}
             
\section{\label{appendix:matched_samples}Matched samples} 
\begin{figure}[ht]
    \centering
\includegraphics[width=0.48\textwidth]{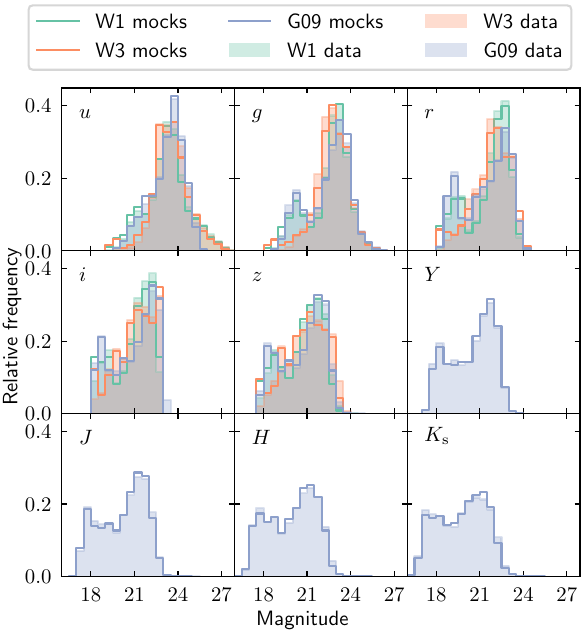}
    \hfill \includegraphics[width=0.48\textwidth]{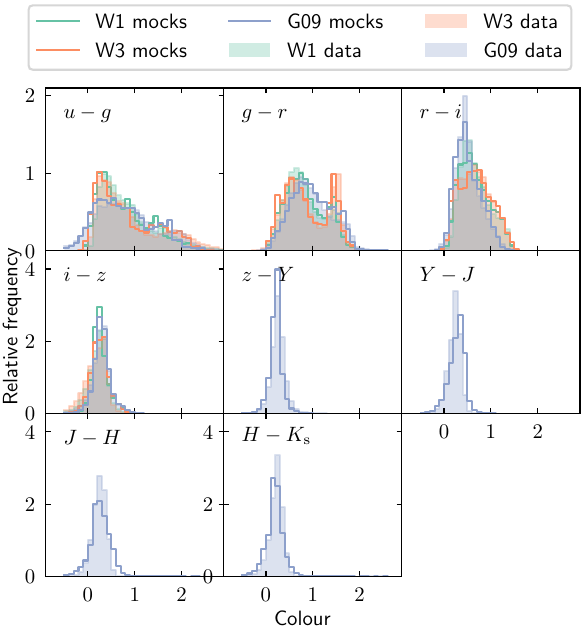}
    \caption{\textit{Top}: Comparison of magnitudes between the spectroscopic validation samples from PAUS data (shaded histograms) and the corresponding mock samples (outlined histograms) in the three PAUS wide fields. \textit{Bottom}: Galaxy colour distributions for the same samples.} \label{fig:matched_sample_distr_plots}
\end{figure}
Spectroscopic redshifts (spec-$z$s) are essential to validate the PAUS photometric redshifts. The validation samples for PAUS are described in detail in \citetalias{navarro-gironesPAUSurveyPhotometric2023}. Given the wide angular separation of the PAUS fields, multiple spectroscopic surveys are combined. We use data from the SDSS DR16 \citep[][]{ ahumadaSixteenthDataRelease2020}, GAMA DR3 \citep{baldryGalaxyMassAssembly2018}, VIPERS \citep[][]{scodeggioVIMOSPublicExtragalactic2018}, DEEP2 Galaxy Redshift Survey \citep{newmanDEEP2GalaxyRedshift2013}, KiDZ-COSMOS compilation from KiDS DR5 \citep{wrightFifthDataRelease2024}, 2-degree Field Galaxy Redshift Survey \citep[2dFGRS,][]{colless2dFGalaxyRedshift2001}, VVDS \citep[][]{lefevreVIMOSVLTDeep2013}, 3-Dimensional Hubble Space Telescope \citep[3D-HST,][]{brammer3DHSTWidefieldGrism2012}, and some smaller numbers of spec-$z$s from other surveys. 
In terms of field coverage, Table\,\ref{tab:specz_samples} describes which surveys cover the individual PAUS fields. 
The distributions of the validation samples as a function of $i_\mathrm{AB}$ and $z_\mathrm{sp}$ are more consistent with the full PAUS catalogue for W1 and W3, whereas the G09 validation sample is more concentrated around $i_\mathrm{AB}=19$, requiring indirect validation through KiDZ-COSMOS. For all fields, the number of spec-$z$ objects drops beyond $z_{\rm sp}\simeq1$.

One part of the analysis showcases the fair comparison between the photo-$z$ statistics of the data and the mocks, which requires a reconstruction of the spec-$z$ validation sample. 
The matching algorithm described in Sect.\,\ref{sec:matched_sample} results in a subsample of simulated galaxies that resembles the spec-$z$ validation sample in terms of redshift and magnitude distributions by design. 
Figure\,\ref{fig:matched_sample_distr_plots} illustrates how well the properties of the spectroscopic data sets agree with the simulated validation samples, shown separated by the filter band (top) or the colour derived from adjacent filters (bottom). For the W1 and W3 fields, the broad bands from CFHTLenS are used, while the G09 field is shown with the nine-band photometry from KiDS. With the \textit{k}NN-matching procedure, we ensure consistent magnitude-dependent statistics for photo-$z$ validation. The mock samples also accurately reproduce the colour distributions of the spectroscopic validation samples, ensuring that the selection function is well captured for photo-$z$ performance assessments.

\begin{table}[h!]
\caption{Main spectroscopic redshift surveys used in the W1, W3, and G09 fields. The first column gives the name of each spectroscopic survey. The second, third, and fourth columns give the number of spectroscopic redshifts in each field.}
\begin{center}
\smallskip
\label{tab:specz_samples}
\smallskip
\begin{tabular}{lrrr}
\hline\hline
\noalign{\vskip 2pt}
Survey & W1    & W3   & G09 \\
\hline
  \noalign{\vskip 3pt}
SDSS         & 5437  & 8018 & 1213 \\ 
GAMA         & 8884  & 0    & 4704 \\ 
VIPERS       & 21\,378 & 0    & 0    \\ 
DEEP2        & 0     & 6969 & 0    \\ 
KiDZ-COSMOS  & 0     & 0 & 11\,854    \\ 
2dFGRS & 2662  & 0    & 0    \\ 
VVDS         & 2216  & 0    & 0    \\ 
3DHST        & 933   & 707  & 0    \\ 
Miscellaneous        & 1193   & 130  & 0    \\ \hline
Total        & 42\,703 & 15\,824 & 17\,771    \\ \hline
\end{tabular}
\end{center}
\end{table}

\section{\label{appendix:default_noise}Default noise in PAUS narrow bands}
\begin{figure*}[h]
    \centering
    \includegraphics[width=\textwidth]{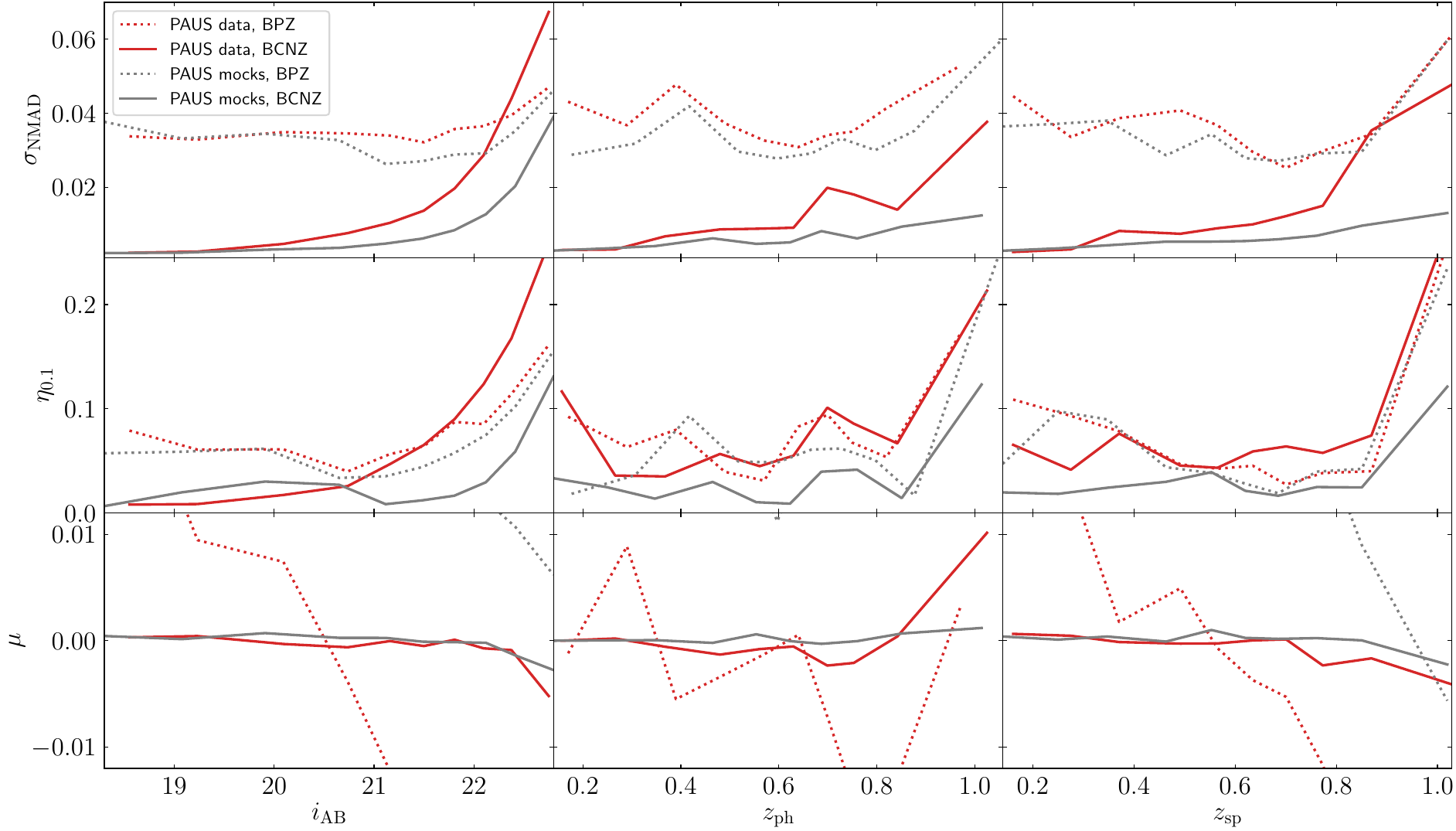}
    \caption{
    Same as Fig.\,\ref{fig:photo-z_statistics_matched}, but now we compare the data (red) and the matched mocks (grey), specifically the \texttt{BCNZ} photo-$z$s (solid lines) and the \texttt{BPZ} photo-$z$s (dotted lines), to analyse the differences and similarities in these statistics. Here again, the \texttt{BPZ} results are obtained using only the external broad-band photometry for galaxies located in the PAUS footprint, while \texttt{BCNZ} additionally uses the PAUS narrow-band data.
    }
    \label{fig:photo-z_statistics_default_matched}
\end{figure*} 
Figure~\ref{fig:photo-z_statistics_default_matched} summarises the photometric redshift performance for the PAUS-like simulations at the baseline noise level, that is, with photometric uncertainties matching those expected for the PAUS survey under its standard observing conditions (see Fig.\,\ref{fig:flux_comp}). We compare two algorithms: \texttt{BCNZ} photo-$z$s (solid lines, 40 narrow bands plus broad bands), and \texttt{BPZ} photo-$z$s (dotted lines, broad bands only). Results are shown both for the simulated matched mock catalogue (blue) and the PAUS data (red). 
In Fig.\,\ref{fig:photo-z_statistics_default_matched}, all the metrics show a stronger correlation with the magnitude than with redshift. When comparing the photometric redshifts obtained from observations (red solid and dotted lines), \texttt{BCNZ} measurements exhibit a strong correlation between $\sigma_\sfont{NMAD}$ and observed magnitude, while the \texttt{BPZ} estimates remain approximately constant across the full magnitude range, as was already shown in \citetalias{navarro-gironesPAUSurveyPhotometric2023}. Overall, \texttt{BCNZ} shows a better performance than \texttt{BPZ} for all the considered metrics, except at the faintest magnitudes ($i_{\rm AB} \gtrsim 22$), where $\sigma_\sfont{NMAD}$ and $\Delta \eta_{\sfont{0.1}}$ are lower for \texttt{BPZ} measurements. 

When comparing the photo-$z$s obtained from the mock and observed data, we obtain similar trends between the metrics but with significant differences. 
For the photo-$z$ scatter, the \texttt{BPZ} mocks show very good agreement with the data across both magnitude and redshift, closely following the observed trends. In contrast, the \texttt{BCNZ} mocks systematically underestimate the scatter, with the discrepancy becoming more pronounced at faint magnitudes.
A similar pattern is observed for the outlier fraction. At bright magnitudes, both \texttt{BPZ} and \texttt{BCNZ} mocks agree well with the data. Toward fainter magnitudes and higher redshifts, the \texttt{BPZ} mocks successfully reproduce the increasing outlier fraction seen in the observations, whereas the \texttt{BCNZ} mocks consistently predict fewer outliers than observed.
For the redshift bias, the \texttt{BCNZ} mocks are in good agreement with the measurements from the data. The \texttt{BPZ} mocks reproduce the overall trends but exhibit a systematic offset, with lower bias values than observed across all magnitude and redshift ranges. This offset is expected to be reduced once the bias correction is applied.

In summary, the \texttt{BPZ} photo-$z$s successfully reproduce the photo-$z$ scatter and outlier fraction observed in the PAUS data. In contrast, the \texttt{BCNZ} estimates systematically show lower scatter and fewer outliers than measured in the data, suggesting that the noise in the narrow bands might be underestimated. As discussed in Sect.~\ref{sec:adding_noise}, this mismatch likely arises from additional sources of photometric uncertainty present in the real observations but not fully captured by the simulations, such as calibration uncertainties, aperture effects, sky subtraction residuals, scattered light, or source blending. A complementary explanation is that the template libraries employed both for the \texttt{BCNZ} photo-$z$ estimation and in the Flagship 2 simulations are closely matched, which leads to more accurate photometric redshift estimates in the mocks.
To mitigate these discrepancies in the photo-$z$ statistics, we apply an empirical rescaling of the photometric noise as described in Sect.~\ref{sec:adding_noise}, effectively broadening the flux uncertainty distributions and bringing the mock photo-$z$ performance into closer agreement with the observations.

\section{\label{appendix:complete_samples} Photometric redshift behaviour in matched and full samples}

\begin{figure*}
    \centering
    \includegraphics[width=\textwidth]{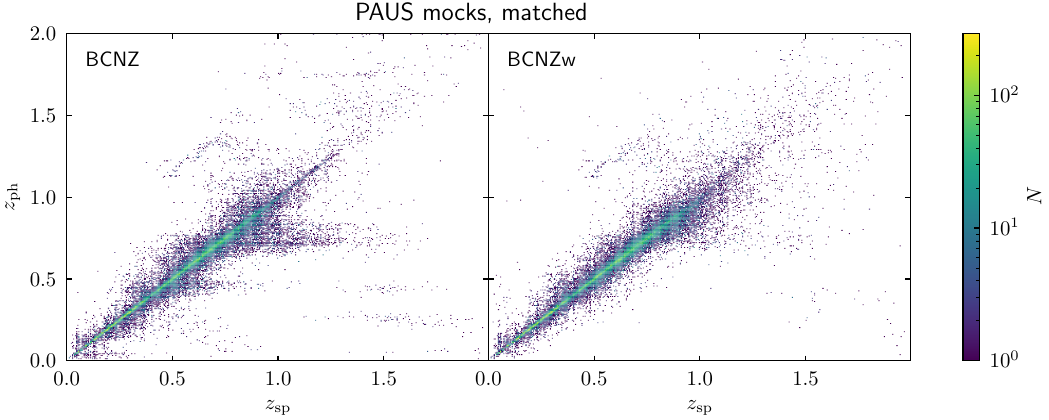}
    \caption{Photometric redshift $z_{\rm ph}$ measured with \texttt{BCNZ} (left) and \texttt{BCNZ}w (right) as a function of the true redshift from the Flagship 2 catalogue for the matched PAUS sample. The colour bar indicates the density of objects.  
    }
    \label{fig:scatter_plot_matched}
\end{figure*}

Figure\,\ref{fig:scatter_plot_matched} shows the relation between the estimated redshift and the true redshift, which is one of the standard diagnostics of photo-$z$ quality. 
At low $z_{\rm sp}$, where we tend to find brighter objects, most galaxies cluster around the one-to-one relation, demonstrating the high precision achievable with narrow-band photo-$z$s. However, at higher redshifts and therefore also fainter magnitudes, the scatter broadens and we can identify locations in this plot with a noticeable fraction of outliers. This hints that there are systematic failures in the photo-$z$ estimation, which could be due to noisy photometry of faint galaxies. 
These structures often trace degeneracies in galaxy SED templates (e.g., confusion between the 4000\,\AA\,\,break and emission line features) and highlight systematic limitations of template-based methods such as \texttt{BCNZ}. 

Moreover, when the observed photometry of galaxies is affected by large flux uncertainties, the likelihood function becomes broad and only weakly constraining. In this regime, the photometric redshift posterior is dominated by the assumed prior (e.g., magnitude or surface-brightness priors). 
As a consequence, the inferred redshifts of noisy or faint galaxies tend to concentrate around redshift regions favoured by the prior or by template degeneracies. 
This phenomenon, also referred to as `redshift focussing' \citep[e.g.,][]{hildebrandtCFHTLenSImprovingQuality2012}, can produce artificial peaks in the estimated redshift distribution $N(z_{\mathrm{ph}})$, even when the true distribution is smooth.  
For \texttt{BCNZ} (see Fig.\,\ref{fig:scatter_plot_matched}, left), we can find these structured patterns showing the effect of redshift focussing at $z_{\rm BCNZ}=0.7$--0.8, which was already demonstrated in 
\citetalias{navarro-gironesPAUSurveyPhotometric2023}. The location of the accumulation point varies depending on the available broad-band photometry. For fields using CFHTLenS photometry, the accumulation occurs near $z_{\rm BCNZ}=0.72$, whereas for the KiDS fields it shifts to $z_{\rm BCNZ}=0.89$. This behaviour is consistent with the broader wavelength coverage of the KiDS+VIKING photometry, including near-infrared bands, which modifies the effective prior constraints. This redshift focussing is partly alleviated for the weighted version of the photo-$z$ (see Fig.\,\ref{fig:scatter_plot_matched}, right).
A study by \citet{hernan-caballeroMiniJPASSurveyMaximising2024} combining the full redshift probability distributions of narrow-band photo-$z$s and broad-band-only photo-$z$s showed similar success. Together, they demonstrate the value of hybrid approaches that exploit both quasi-spectroscopic information from narrow bands and broad-band colour priors.

\begin{figure*}
    \centering
\includegraphics[width=\textwidth]{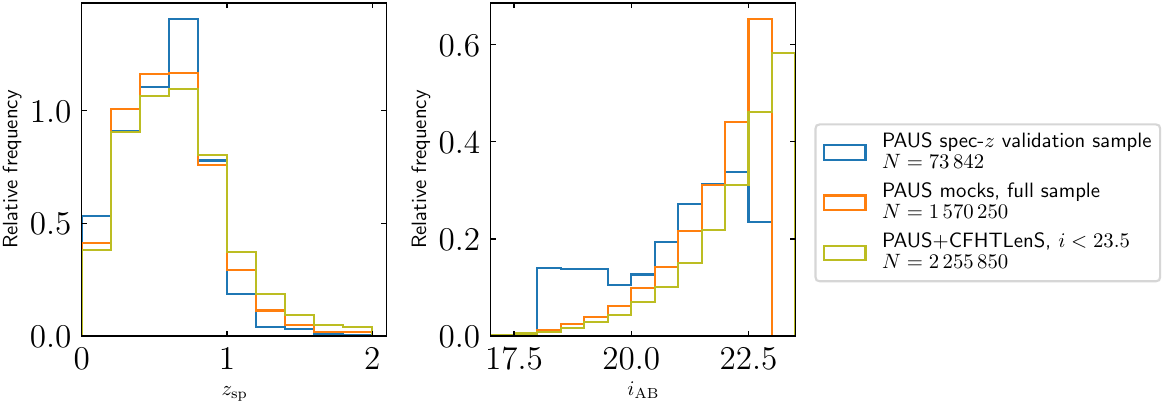}
    \caption{Redshift and $i_{\rm AB}$ distribution of the spectroscopic validation sample, the full sample of PAUS and the extended deeper version of the simulated sample.
    }
\label{fig:redshift_distr_i_full}
\end{figure*}

While for the PAUS data, we are limited to the spec-$z$ validation sample for testing the performance of the PAUS photo-$z$s, the mocks have the true redshift for all galaxies, which is a much larger and fainter sample, as can be seen in Fig.\,\ref{fig:redshift_distr_i_full}.
\begin{figure*}
    \centering
    \includegraphics[width=\textwidth]{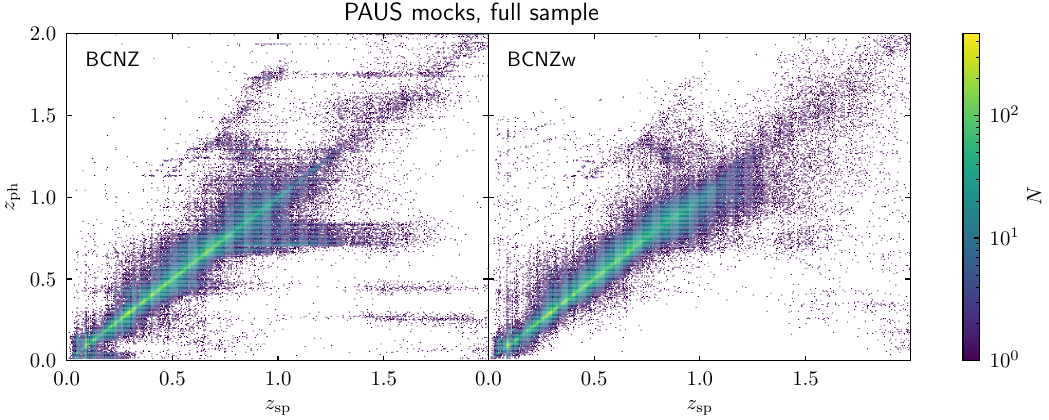}
    \caption{Same as Fig.\,\ref{fig:scatter_plot_matched}, but this time for the full PAUS mock sample with $i_{\rm AB} < 23$. }
    \label{fig:scatter_plot_all}
\end{figure*}
Compared to Fig.\,\ref{fig:scatter_plot_matched}, there are naturally more objects at high redshifts in Fig.\,\ref{fig:scatter_plot_all} because of the fainter galaxy population. The fainter objects also have more noise in the photometry and therefore, the scatter and outlier numbers are visibly increased. The weighted photo-$z$ estimate (right) reduces some of the systematic structures of the $z_{\rm ph}$--$z_{\rm sp}$ plane that are visible for \texttt{BCNZ} (left).
Also shown in Fig.\,\ref{fig:redshift_distr_i_full} is the sample PAUS+CFHTLenS for $i<23.5$ which is presented in Sect.\,\ref{sec:deep_paus}. At the faint end, a large fraction of galaxies are included in the sample relative to the full PAUS mock sample for $i<23$.

\section{\label{appendix:forecasts}Forecasts with PAUS deep and \Euclid}
\begin{figure*}[h]
    \centering
    \includegraphics[width=\textwidth]{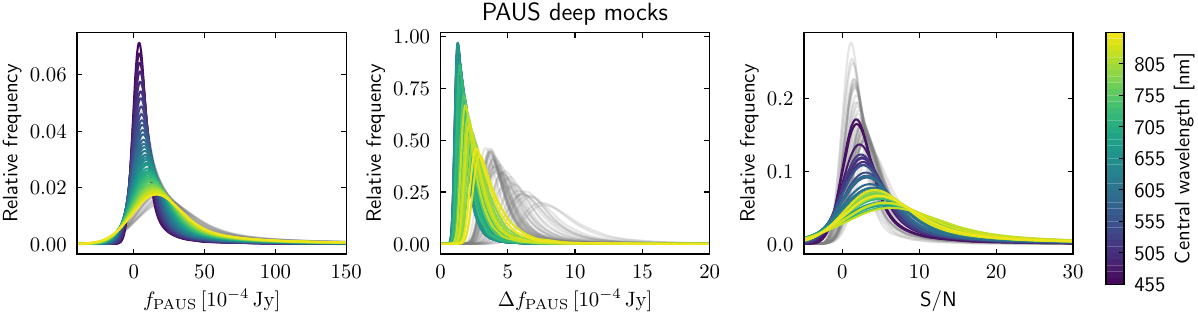}
    \caption{Distribution of flux (left), flux uncertainty (centre), and resulting S/N (right) for the deeper simulated version of PAUS. For comparison, the baseline PAUS data from Fig.\,\ref{fig:flux_comp} are shown as grey lines. The deeper version contains many more faint objects, and the combination of improved seeing (PSF FWHM) and deeper limiting magnitudes (see Fig.\,\ref{fig:mag_limits_psf_data}) suppresses flux uncertainties, leading to a systematically higher S/N.}
    \label{fig:flux_deep}
\end{figure*}

Figure~\ref{fig:flux_deep} illustrates the photometric properties of the deeper simulated PAUS configuration introduced in Sect.~\ref{sec:deep_paus}. Shown are the distributions of flux, flux uncertainty, and S/N for the narrow-band filters, with the baseline PAUS data overplotted for reference. The deeper configuration contains a larger population of faint objects, while simultaneously exhibiting reduced flux uncertainties. In our forecast, we attribute this improvement to the combination of increased limiting magnitudes and improved seeing, which enhances the effective S/N. Consequently, the S/N distributions broaden systematically to higher values across all narrow bands.

At the magnitude limit of the forecast sample ($i_{\rm AB}=23.5$), the median narrow-band S/N is approximately 1.1 for the standard PAUS configuration and 1.3 for the deeper PAUS setup, while the broad-band photometry typically reaches median S/N values of $\sim10$--20. 
For galaxies with high narrow-band S/N, all survey configurations achieve similarly low photo-$z$ scatter. The main difference between the configurations is therefore the fraction of galaxies that remain above the S/N threshold at which narrow-band spectral features can still reliably constrain the redshift. 
At low narrow-band S/N, the additional \Euclid near-infrared photometry reduces the photo-$z$ scatter, outlier fraction, and bias by improving continuum constraints and alleviating template degeneracies. 

\begin{figure*}
    \centering
    \includegraphics[width=\textwidth]{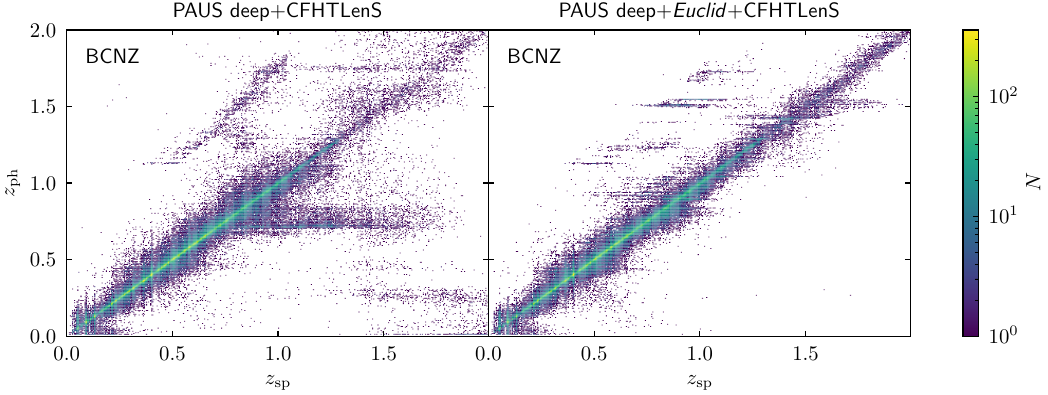}
    \caption{The \texttt{BCNZ} photo-$z$s as a function of the true redshift from the Flagship 2 catalogue for the simulated PAUS deep+CFHTLenS sample (left) and when also adding the \Euclid photometry (right). Both samples have $i_{\rm AB} <23.5$.}
\label{fig:scatter_plot_pausdeep+euclid}
\label{LastPage}
\end{figure*}

Figure~\ref{fig:scatter_plot_pausdeep+euclid} illustrates the impact of survey depth and complementary broad-band photometry on the photo-$z$ performance in the mock forecasts discussed in Sect.~\ref{sec:deeper_paus}. We show \texttt{BCNZ} photo-$z$s as a function of the true redshift for galaxies with $i_{\rm AB}<23.5$.
Figure~\ref{fig:scatter_plot_pausdeep+euclid} presents the deeper PAUS forecast combined with CFHTLenS broad-band data (left), and the same setup with the addition of \Euclid near-infrared photometry (right). 
The deeper narrow-band data (left panel) results in improvements of the photo-$z$ statistics (see Fig.\,\ref{fig:photoz_statistics_euclid}), but does not show visible changes in the structure of the $z_{\rm ph}$–$z_{\rm true}$ scatter plots when comparing to the left panel of Fig.~\ref{fig:scatter_plot_all}. The effect of redshift focussing is now more pronounced around $z_{\rm ph}=0.72$, since we are only using the CFHTLenS broad bands (no KiDS photometry). 
However, including \Euclid photometry (right panel) significantly reduces the concentration of systematic outliers.  Moreover, the overall scatter is reduced by extending the wavelength coverage into the near-infrared as it improves the constraints on the galaxy SEDs. The additional broad- and near-infrared photometry reduces the dominance of the prior in cases where the PAUS narrow-band measurements alone have low signal-to-noise. In particular, the extended wavelength baseline helps to break degeneracies between galaxy templates and redshift solutions, leading to a broader distribution of posterior peaks instead of an artificial accumulation at the preferred prior redshift. Consequently, the redshift-focussing feature becomes strongly suppressed when including the \Euclid-like photometry.
This figure provides a visual illustration of the trends quantified in Sect.~\ref{sec:deeper_paus}: deeper narrow-band data primarily improve photo-$z$ precision, while the addition of \Euclid photometry is especially effective at mitigating systematic failures and redshift degeneracies, resulting in more robust photo-$z$ estimates for faint and high-redshift galaxies.

\end{appendix}
\label{LastPage}
\end{document}